\documentclass[%
 reprint,
 amsmath,amssymb,
 aps,
]{revtex4-1}

\usepackage{graphicx}% Include figure files
\usepackage{dcolumn}% Align table columns on decimal point
\usepackage{bm}% bold math
\usepackage{pst-all}
\input epsf
\usepackage{epic}
\usepackage{pstricks}
\usepackage{epsfig}
\usepackage{pst-grad} % For gradients
\usepackage{pst-plot} % For axes
\usepackage[space]{grffile} % For spaces in paths
\usepackage{etoolbox} % For spaces in paths
\makeatletter % For spaces in paths
\patchcmd\Gread@eps{\@inputcheck#1 }{\@inputcheck"#1"\relax}{}{}
\makeatother

\begin{document}

\title{Time transfer functions without enhanced terms in stationary spacetimes \\
Application to an isolated, axisymmetric spinning body}
% Force line breaks with \\

\author{P. Teyssandier}
\email{Pierre.Teyssandier@obspm.fr}
\affiliation{SYRTE, Observatoire de Paris, PSL Research University, CNRS, Sorbonne Universit\'es, \\
 UPMC Univ. Paris 06, LNE, 61 avenue de l'Observatoire, F-75014 Paris, France}

\date{\today}% It is always \today, today,

\begin{abstract}
We develop a new perturbation method for determining a class of time transfer functions in a stationary spacetime when its metric is a small deformation of a background metric for which the time transfer functions are known in a closed form. The perturbation terms are expressed as line integrals along the null geodesic paths of the background metric. Unlike what happens with the other procedures proposed until now, the time transfer functions obtained in this way are completely free of unbounded terms and do not generate any enhancement in the light travel time. Our procedure proves to be very efficient when the background metric is a linearized 
Schwarzschild-like metric. Its application to an isolated body slowly rotating about an axis of symmetry leads to integrals which can be  calculated with any symbolic computer program.
Explicit  expressions are obtained for the mass dipole and quadrupole moments and for the leading gravitomagnetic term induced by the spin of the body. A brief numerical discussion is given for 
the 2002 Cassini experiment.
   
\begin{description}
\item[PACS numbers]
04.25.-g, 04.80.Cc, 95.10.Jk, 95.30.Sf
\end{description}
\end{abstract}

$ $

\pacs{04.25.-g, 04.80.Cc, 95.10.Jk, 95.30.Sf}% PACS, the Physics and Astronomy
\maketitle

%\tableofcontents

%%%%%%%%%%%%%%%%%%%%%%%%%%%%%%%%%%%%%%%%%%%%%%%%
\section{Introduction} \label{sec:intro}
%%%%%%%%%%%%%%%%%%%%%%%%%%%%%%%%%%%%%%%%%%%%%%%%

Determining the travel time of an electromagnetic signal as a function only involving the spatial position of the emitter, the spatial position of the receiver and the instant of arrival of the signal is a problem of crucial importance in metric theories of gravity. Indeed, knowing such a 
function -- called a reception time transfer function (TTF) according to a terminology specified 
in \cite{Linet:2002,Leponcin:2004} -- makes it possible to model the relativistic time delay 
and/or the Doppler observables as measured in space missions (see, e.g., \cite{Bertotti:2008,Ashby:2010,Hees1:2012,Hees2:2014,Hees3:2014}) or in Very Long Baseline Interferometry observations \cite{Leponcin:2016,Soffel:2017}. It also enables one to 
calculate the propagation direction of a light ray at the points of emission and observation \cite{Leponcin:2004,Leponcin:2008,Teyssandier:2012}, thus justifying the relevancy of the notion 
of TTF in relativistic astrometry \cite{Bertone:2014,Bertone:2017}. Moreover, the TTF 
formalism turns out to be effective in studying the phase function associated with a light ray 
\cite{Qin:2017} and the propagation of light rays in nondispersive isotropic moving media \cite{Bourgoin:2020,Bourgoin:2021,Bourgoin:2022}.

The explicit calculation of TTFs with the accuracy required by current space missions and/or  
relativistic astrometry has given rise to a large body of works. Two kinds of procedures have 
been systematically investigated, each based on the assumption that the TTFs involved in the observations made in the Solar system can be represented by a general post-Newtonian or 
post-Minkowskian expansion. The first kind of procedures consists in determining the light rays 
by integration of the null geodesic equations by approximate asymptotical methods (see, e.g., \cite{Klioner:1992,Kopeikin:1999,Kopeikin:2006,Klioner:2010,Korobkov:2014,Zschocke:2015,
Zschocke:2018,Zschocke1:2022} and Refs. therein). The second kind of approaches avoids the integration of the geodesic equations either by using Synge's world function \cite{Linet:2002,Leponcin:2004} or by integrating an eikonal equation \cite{Ashby:2010,Teyssandier:2008}. These different procedures lead to equivalent formulas \cite{Zschocke:2014}. However, they suffer from two major shortcomings. First, they allow to determine just one TTF, even in configurations where a gravitational lensing phenomena may occur \cite{Zschocke:2011}. Second, the TTF so far obtained for a large class of parametrized static spherically symmetric metrics contain enhanced terms, i.e. terms which become infinite when the emission and reception points tend to be diametrically opposed \cite{Teyssandier:2008,Ashby:2010,Klioner:2010,Linet:2013,Teyssandier:2014}. The same issue 
arises when the multipole moments or/and the motion of the source(s) of the field are taken into account (see, e.g., \cite{Kopeikin:1997,Kopeikin:2002,Linet:2002,Ciufolini:2003,Leponcin:2008,Korobkov:2014,Hees3:2014,
Soffel:2015,Zschocke:2018,Zschocke2:2022}). 

Clearly, we must abandon the assumption that a TTF can be represented by a post-post-Newtonian 
or a generalized post-Minkowskian expansion regardless of the positions of the emitter and the 
receiver. This crucial point is confirmed in \cite{Linet:2016}, where all the TTFs of a Schwarzschild-like metric are obtained in an exact closed-form within the linearized, weak-field approximation. These functions are bounded everywhere, but it is proven that none of 
them can be represented by a convergent post-Minkowskian expansion when the emitter and the receiver are sufficiently close to a superior conjunction. 

The aim of the present paper is to extend the approach developed in \cite{Linet:2016} to the case where the metric is a weak stationary perturbation of a background metric whose null geodesics 
and their associated TTFs are explicitly known in an analytical closed form. The perturbation 
terms in each TTF are obtained as line integrals evaluated along a null geodesic path of the background metric. The case of a weak stationary perturbation of a linearized Schwarzschild-like metric is treated in detail. We apply these results to light rays propagating in the field 
of an isolated body slowly rotating about an axis of symetry. We show that the perturbations 
in the propagation time of light rays due to the mass and spin multipoles can be calculated with any symbolic computer program. Explicit expressions are derived for the dipole and quadrupole 
mass moments, as well as for the dominant gravitomagnetic term due to the intrinsic angular momentum of the body. All the expressions are bounded when the positions of the emitter and the receiver tend to be diametrically opposite. 

The applications of our results to the deflection of light and the gravitational lensing require lengthy calculations which are beyond the scope of the present work. So we restrict ourselves to comparing the new formulas with those previously used to determine the post-Newtonian parameter $\gamma$ from data collected during the Cassini mission.

In this paper, $G$ is the Newtonian gravitational constant and $c$ is the speed of light in a vacuum. Greek indices run from 0 to 3, and Latin indices run from 1 to 3. Any bold italic letter stands for an ordered triple: we note, e.g., $\bm{a}=(a^1,a^2,a^3)=(a^i)$. All the 
triples are formally regarded as vectors of a three-dimensional Euclidean space. Given two 
triples $\bm{a}$ and $\bm{b}$, $\bm{a}.\bm{b}$ and $\bm{a}\times\bm{b}$ denote their usual scalar product and exterior product, respectively. The notation $\vert\bm{a}\vert$ stands for the Euclidean norm of $\bm{a}$. The Einstein summation convention is used for both types of repeated indices, whatever their respective positions. 

%%%%%%%%%%%%%%%%%%%%%%%%%%%%%%%%%%%%%%%%%%%%%%%%%%%%%%%%%%%%%%%%%%%%%%%%%%%%%%%%%%%%%%%
\section{Time transfer functions in stationary spacetimes} \label{sec:dTTF}
%%%%%%%%%%%%%%%%%%%%%%%%%%%%%%%%%%%%%%%%%%%%%%%%%%%%%%%%%%%%%%%%%%%%%%%%%%%%%%%%%%%%%%%

Throughout this work, the physical metric $g$ is supposed to be stationary. The signature 
adopted for $g$ is $(+,-,-,-)$. Spacetime is assumed to be covered by a global coordinate system $(x^0=ct,\bm{x})$, $t$ being a time-like coordinate and $\bm{x}$ a triple 
of quasi-Cartesian spatial coordinates: $\bm{x}=(x^i)$. This coordinate system is chosen in 
such a way that the components of the metric do not depend on $x^0$. The light signals are 
supposed to propagate through a vacuum, so the light rays are null geodesic paths of the metric $g$. 

Let $\mathcal{G}^{[\sigma]}(\bm{x}_A,x^0_B,\bm{x}_B)$ be a generic member of the 
family of light rays starting from points-events having a given spatial position $\bm{x}_A$ 
and arriving at the point-event $(x_B^0,\bm{x}_B)$, the brackets around the index 
$\sigma$ being used to avoid any confusion with an exponent or a tensorial index. Let us denote 
by $(x^{0}_A)^{[\sigma]}$ the value of $x^0$ when the light ray $\mathcal{G}^{[\sigma]}(\bm{x}_A,x^0_B,\bm{x}_B)$ is emitted. Owing to the stationary character 
of spacetime, the light travel time $[x^{0}_B-(x^{0}_A)^{[\sigma]}]/c$ corresponding to 
this ray depends only on the spatial positions $\bm{x}_A$ and $\bm{x}_B$. We can 
therefore write
\begin{equation} \label{TTFd}
x^{0}_B-(x^{0}_A)^{[\sigma]}=c\mathcal{T}^{[\sigma]}(\bm{x}_A,\bm{x}_B),
\end{equation}  
where $\mathcal{T}^{[\sigma]}$ is the reception time transfer function associated with the light ray $\mathcal{G}^{[\sigma]}(\bm{x}_A,x^0_B,\bm{x}_B)$. It may be noted that each path $\mathcal{G}^{[\sigma]}(\bm{x}_A,x^0_B,\bm{x}_B)$ can be deduced from $\mathcal{G}^{[\sigma]}(\bm{x}_A,0,\bm{x}_B)$ by a timelike translation. So, in what follows, any light ray of type $\sigma$ starting from $\bm{x}_A$ and arriving at 
$\bm{x}_B$ will be simply denoted by $\mathcal{G}^{[\sigma]}(\bm{x}_A,\bm{x}_B)$. Moreover, the term ``reception" will be systematically omitted.

%%%%%%%%%%%%%%%%%%%%%%%%%%%%%%%%%%%%%%%%%%%%%%%%%%%%%%%%%%%%%%%%%%%%%%%%%%%%%%%%%%%%%%%%%%%%%%%
\section{Perturbation approach of the TTF formalism in stationary spacetimes} \label{sec:pert}
%%%%%%%%%%%%%%%%%%%%%%%%%%%%%%%%%%%%%%%%%%%%%%%%%%%%%%%%%%%%%%%%%%%%%%%%%%%%%%%%%%%%%%%%%%%%%%%

Let us now enunciate the other general assumptions underlying this paper.

1. The metric $g$ depends on a family of small dimensionless parameters $\{\epsilon_{a}, a\in \mathcal{P}\}$ describing the departure of this metric from a background stationary metric 
denoted by $\gamma$; $\mathcal{P}$ may be an infinite set. 

2. The components $g_{\mu\nu}$ of the metric can be decomposed as
\begin{equation} \label{ggh}
g_{\mu\nu}(\bm{x};\mathcal{P})=\gamma_{\mu\nu}(\bm{x})+h_{\mu\nu}(\bm{x};\mathcal{P}),
\end{equation}
where the perturbation terms $h_{\mu\nu}$ are assumed to be represented with a sufficient 
accuracy by an expansion as follows
\begin{equation} \label{expgbe}
h_{\mu\nu}(\bm{x};\mathcal{P})=\sum_{a \in \mathcal{P}}h_{\mu\nu}^a(\bm{x})+O(\epsilon^2),
\end{equation}
each $h_{\mu\nu}^a$ being the first-order perturbation due to $\epsilon_{a}$ and $O(\epsilon^2)$ standing for all the terms of order $\epsilon_{a}\epsilon_{b}$, with $(a,b)\in\mathcal{P}\times\mathcal{P}$. Then, the contravariant components of the metric tensor 
may be written as
\begin{equation} 
g^{\mu\nu}(\bm{x};\mathcal{P})=\gamma^{\mu\nu}(\bm{x})+\sum_{a\in\mathcal{P}}k^{\mu\nu}_{a}(\bm{x})+O(\epsilon^2),
\label{expghe}
\end{equation}
where the $\gamma^{\mu\nu}$ are coefficients satisfying the equations $\gamma_{\mu\rho}\gamma^{\mu\nu}=\delta_{\rho}^{\nu} $ and the quantities $k^{\mu\nu}_{a}$ are 
given by
\begin{equation}
k^{\mu\nu}_{a}(\bm{x})=-\gamma^{\mu\alpha}(\bm{x})\gamma^{\nu\beta}(\bm{x})h_{\alpha\beta}^{a}(\bm{x}). 
\label{kmn}
\end{equation} 

We systematically use greek letters to denote the mathematical objects which 
only depend on the background metric $\gamma$. We also use the zeroth-order appellation for these objects. No confusion is possible between the background metric and the post-Newtonian parameter $\gamma$. 

3. In agreement with a notation already specified in Sect. \ref{sec:dTTF}, let us denote by $\Gamma^{[\sigma]}(\bm{x}_A,\bm{x}_B)$ a generic member of the family of null geodesic paths 
of the background metric $\gamma$ starting from the spatial position $\bm{x}_A$ and arriving at 
the point-event $(x^0_B,\bm{x}_B)$, and by $T^{[\sigma]}(\bm{x}_A,\bm{x}_B)$ the corresponding zeroth-order time transfer function. It is assumed that whatever the value of the parameter $\sigma$, there exists a null geodesic $\mathcal{G}^{[\sigma]}(\bm{x}_A,\bm{x}_B;\mathcal{P})$ 
of the metric $g$ such that the corresponding TTF $\mathcal{T}^{[\sigma]}(\bm{x}_A,\bm{x}_B;\mathcal{P})$ can be written as 
\begin{eqnarray}
&&\mathcal{T}^{[\sigma]}(\bm{x}_A, \bm{x}_B;\mathcal{P})=T^{[\sigma]}(\bm{x}_A, \bm{x}_B)+\sum_{a\in\mathcal{P}}\Delta\mathcal{T}^{[\sigma]}_a(\bm{x}_A,\bm{x}_B) \nonumber \\
&&\qquad\qquad\qquad\quad\quad+O(\epsilon^2),
\label{expTe}
\end{eqnarray} 
where each $\Delta\mathcal{T}^{[\sigma]}_a$ denotes the first-order perturbation term due to the 
small parameter $\epsilon_a$.

To determine the perturbation functions $\Delta\mathcal{T}^{[\sigma]}_a$, we use the property of each function $\mathcal{T}^{[\sigma]}$ to obey an eikonal equation which can be written as \cite{Bel:1994,Teyssandier:2008}
\begin{widetext}
\begin{equation}
g^{00}(\bm{x};\mathcal{P})-2g^{0i}(\bm{x};\mathcal{P})\frac{c\partial \mathcal{T}^{[\sigma]}(\bm{x}_A, \bm{x};\mathcal{P})}{\partial x^i}+g^{ij}(\bm{x};\mathcal{P})\frac{c\partial \mathcal{T}^{[\sigma]}(\bm{x}_A,\bm{x};\mathcal{P})}{\partial x^i}\frac{c\partial \mathcal{T}^{[\sigma]}(\bm{x}_A,\bm{x};\mathcal{P} )}{\partial x^j}=0.
\label{eik}
\end{equation}

Substituting the expansions (\ref{expghe}) and (\ref{expTe}) into Eq. (\ref{eik}), and then retaining only the terms of orders 0 and 1 with respect to the parameters $\epsilon_a$, we get 
the system of equations
\begin{eqnarray}
&&\gamma^{00}(\bm{x})-2\gamma^{0i}(\bm{x})\frac{c\partial T^{[\sigma]}(\bm{x}_A,\bm{x})}{\partial x^i}+\gamma^{ij}(\bm{x})\frac{c\partial T^{[\sigma]}(\bm{x}_A,\bm{x})}{\partial x^i}\frac{c\partial T^{[\sigma]}(\bm{x}_A,\bm{x} )}{\partial x^j}=0, \label{eikT0}\\
&&k_{a}^{00}(\bm{x})-2k_{a}^{0i}(\bm{x})\frac{c\partial T^{[\sigma]}(\bm{x}_A,\bm{x})}{\partial x^i}+k_{a}^{ij}(\bm{x})\frac{c\partial T^{[\sigma]}(\bm{x}_A,\bm{x})}{\partial x^i}\frac{c\partial T^{[\sigma]}(\bm{x}_A,\bm{x})}{\partial x^j} \nonumber \\
&&+2\left[-\gamma^{0j}(\bm{x})+\gamma^{ij}(\bm{x})\frac{c\partial T^{[\sigma]}(\bm{x}_A, \bm{x})}{\partial x^i}\right]\frac{c\partial\Delta\mathcal{T}^{[\sigma]}_{a}(\bm{x}_A,\bm{x})}{\partial x^j}=0.\label{eikT1}
\end{eqnarray}
\end{widetext}
Of course, Eq. (\ref{eikT0}) is the eikonal equation associated with the background metric $\gamma$. 

When $\bm{x}$ is constrained to vary along the zeroth-order light ray $\Gamma^{[\sigma]}(\bm{x}_A,\bm{x}_B)$, Eq. (\ref{eikT1}) can be written in the form of an ordinary first-order differential equation governing the perturbation term $\Delta\mathcal{T}^{[\sigma]}_{a}$. Indeed, let 
\begin{equation} \label{xip}
x^0=\xi^0_{[\sigma]}(v),\quad\bm{x}=\bm{\xi}_{[\sigma]}(v)=(\xi^i_{[\sigma]}(v))            
\end{equation}
be a system of parametric equations of $\Gamma^{[\sigma]}(\bm{x}_A,\bm{x}_B)$, $v$ being 
an arbitrarily chosen parameter. Put
\begin{equation} \label{tgv}
\lambda_{\alpha}^{[\sigma]}(v)=\gamma_{\alpha\mu}(\bm{\xi}_{[\sigma]}(v))\dot{\xi}_{[\sigma]}^{\mu}(v),
\end{equation}
where $\dot{\xi}_{[\sigma]}^{\mu}\!\!=\!\!d\xi_{[\sigma]}^{\mu}/dv$.$\!$ Since the quantities $\lambda_{\alpha}^{[\sigma]}(v)$ are the covariant components of the vector tangent to $\Gamma^{[\sigma]}(\bm{x}_A,\bm{x}_B)$ determined by the background metric $\gamma_{\mu\nu}$, the partial derivatives of $T^{[\sigma]}(\bm{x}_A,\bm{x})$ with 
respect to $x^i$ at point $\bm{\xi}_{[\sigma]}(v)$ satisfy the relations (see Eq. (40) in \cite{Leponcin:2004})
\begin{equation} \label{ki}
\left[\frac{c \partial T^{[\sigma]}(\bm{x}_A,\bm{x})}{\partial x^i}\right]_{\bm{x}=\bm{\xi}_{[\sigma]}(v)}=-\frac{\lambda^{[\sigma]}_i(v)}{\lambda^{[\sigma]}_0(v)}.
\end{equation}
Contracting Eq. (\ref{ki}) by $\gamma^{ij}$ and using the relation 
\begin{equation} \label{glx}
\gamma^{\alpha j}\lambda_{\alpha}^{[\sigma]}(v)=\dot{\xi}_{[\sigma]}^{j}(v),
\end{equation}
it is easily seen that 
\begin{equation} \label{gdT0}
\left[-\gamma^{0j}+\gamma^{ij}\frac{c \partial T^{[\sigma]}(\bm{x}_A,\bm{x})}{\partial x^i}\right]_{\bm{x}=\bm{\xi}_{[\sigma]}(v)}=-\frac{\dot{\xi}_{[\sigma]}^{j}(v)}{\lambda_0^{[\sigma]}(v)}.
\end{equation}
Then, using Eq. (\ref{gdT0}) and noting that 
\begin{eqnarray} 
&&\dot{\xi}_{[\sigma]}^{j}(v)\frac{c\partial\Delta\mathcal{T}^{[\sigma]}_{a}(\bm{x}_A,\bm{\xi}_{[\sigma]}(v))}{\partial x^j} \nonumber \\
&&\qquad\qquad\qquad=\frac{cd\Delta\mathcal{T}^{[\sigma]}_{a}(\bm{x}_A,\bm{\xi}_{[\sigma]}(v))}{dv}, \label{dTa}
\end{eqnarray}
it appears that Eq. (\ref{eikT1}) turns into an ordinary differential equation whose solution is
\begin{equation} \label{DT1}
c\Delta\mathcal{T}^{[\sigma]}_{a}(\bm{x}_A,\bm{x}_B)=\int_{v_A}^{v_B}\!\lambda_0^{[\sigma]}(v)
\mathcal{H}^{[\sigma]}_{a}(\bm{x}_A,\bm{\xi}_{[\sigma]}(v))dv,
\end{equation}
where $v_A$ and $v_B$ are respectively the values of $v$ at points $\bm{x}_A$ and $\bm{x}_B$, and $\mathcal{H}^{[\sigma]}_{a}(\bm{x}_A,\bm{x})$ is given by
\begin{widetext}
\begin{equation} 
\mathcal{H}^{[\sigma]}_{a}(\bm{x}_A,\bm{x})=\frac{1}{2}k_{a}^{00}(\bm{x})-k_{a}^{0i}(\bm{x})\frac{c \partial T^{[\sigma]}(\bm{x}_A,\bm{x})}{\partial x^i} +\frac{1}{2}k_{a}^{ij}(\bm{x})\frac{c \partial T^{[\sigma]}(\bm{x}_A,\bm{x})}{\partial x^i}\frac{c \partial T^{[\sigma]}(\bm{x}_A,\bm{x})}{\partial x^j}.
\label{dI1}
\end{equation}
\end{widetext}

It follows from Eqs. (\ref{tgv}), (\ref{ki}) and (\ref{dI1}) that determining the perturbation 
term $c\Delta\mathcal{T}^{[\sigma]}_{a}$ comes down to calculating a line integral along $\Gamma^{\sigma}(\bm{x}_A,\bm{x}_B)$, the integrand being expressed as a functional only involving the metric components $\gamma_{\mu\nu}$, the functions $\xi^{\mu}(v)$ and their derivatives $\dot{\xi}_{[\sigma]}^{\mu}(v)$.  
 
Omitting the argument $v$ for the sake of brevity, we have  
\begin{equation} \label{lam0}
\lambda_0^{[\sigma]}=\gamma_{00}\dot{\xi}_{[\sigma]}^0+\gamma_{0k}\dot{\xi}_{[\sigma]}^k.
\end{equation}
Squaring each side of Eq. (\ref{lam0}) and taking into account that $\gamma_{\mu\nu}
\dot{\xi}^{\mu}\dot{\xi}^{\nu}=0$ along $\Gamma^{[\sigma]}(\bm{x}_A,\bm{x}_B)$, it may 
be seen that
\begin{equation} \label{dx0dxi}
\lambda_0^{[\sigma]}=\pm\sqrt{\gamma_{00}
\gamma^{\,\ast}_{ij}\dot{\xi}_{[\sigma]}^{i}\dot{\xi}_{[\sigma]}^{j}},
\end{equation} 
the quantities $\gamma^{\,\ast}_{ij}$ being defined as
\begin{equation} 
\gamma^{\,\ast}_{ij}=-\gamma_{ij}+\frac{\gamma_{0i}\gamma_{0j}}{\gamma_{00}}.
\label{gbar}
\end{equation}
Choosing the parameter $v$ so as to have the positive determination of the right-hand side of Eq. (\ref{dx0dxi}), it is immediately seen that Eq. (\ref{DT1}) can be written in the form
\begin{eqnarray} 
&&c\Delta\mathcal{T}^{[\sigma]}_{a}(\bm{x}_A,\bm{x}_B) \nonumber \\
&&=\int_{v_A}^{v_B}\sqrt{\gamma_{00}(\bm{\xi}_{[\sigma]}(v))}\mathcal{H}^{[\sigma]}_{a}(\bm{x}_A,\bm{\xi}_{[\sigma]}(v))d\ell_{[\sigma]},
\label{T1a}
\end{eqnarray}
where $d\ell_{[\sigma]}$ is the infinitesimal distance along $\Gamma^{[\sigma]}(\bm{x}_A,\bm{x}_B)$ defined as
\begin{equation}
d\ell_{[\sigma]}=\sqrt{\gamma^{\,\ast}_{ij}(\bm{\xi}_{[\sigma]}(v))\dot{\xi}_{[\sigma]}^i
\dot{\xi}_{[\sigma]}^j} dv.
\label{dl}
\end{equation}

Since the null geodesics of two conformal metrics are coincident (see, e.g., \cite{Joshi:2007}), 
our problem can be reduced to the calculation of the TTFs for a metric such that 
\begin{equation} \label{g000}
g_{00}=1 
\end{equation}
holds at any point. Accordingly, the corresponding background metric can always be chosen so that 
\begin{equation} \label{ga00}
\gamma_{00}=1
\end{equation}
and the perturbation terms $h^a_{00}$ can be assumed to satisfy the relations
\begin{equation} \label{ha00}
h^a_{00}=0
\end{equation}
for any $a$. This remark will be systematically used in the next sections.

%%%%%%%%%%%%%%%%%%%%%%%%%%%%%%%%%%%%%%%%%%%%%%%%%%%%%%%%%%%%%%%%%%%%%%%%%%%%%%%%%%%%%%%
\section{Application to perturbed Schwarzschild-like spacetimes} \label{sec:pSchw}
%%%%%%%%%%%%%%%%%%%%%%%%%%%%%%%%%%%%%%%%%%%%%%%%%%%%%%%%%%%%%%%%%%%%%%%%%%%%%%%%%%%%%%%

The procedure outlined in the previous section can be applied when the physical metric is 
a weak stationary perturbation of a static, spherically symmetric metric corresponding to 
the field of a central mass $M$. We put $m=GM/c^2$. Choosing appropriate quasi-Cartesian coordinates $(x^0,\bm{x})$ and putting $r=\vert\bm{x}\vert$, the physical $ds^2$ can be 
considered as conformally related to a $d\bar{s}^2$ written in the form
\begin{eqnarray} 
&&d\bar{s}^2=(dx^0)^2+2\bar{h}_{0i}(\bm{x};\mathcal{P})dx^0dx^i\nonumber \\
&&\qquad\;\;\;-\left[\mathcal{U}(r)\delta_{ij}-\bar{h}_{ij}(\bm{x};\mathcal{P})\right]dx^idx^j, \label{bds}
\end{eqnarray} 
where 
\begin{equation} \label{cU}
\mathcal{U}(r)=1+2\kappa_1\frac{m}{r}+2\sum_{n=2}^{\infty}\kappa_n\frac{m^n}{r^n}
\end{equation}
and the quantities $\bar{h}_{\mu i}(\bm{x};\mathcal{P})$ are assumed to be small perturbations 
developable into series analogous to (\ref{expgbe}), the parameters $\epsilon_a$ 
characterizing here the sphericity defect of the central body. The constants $\kappa_n$ are 
some post-Newtonian parameters. In particular, $\kappa_1$ is linked to the Eddington parameter $\gamma$ by the relation  
\begin{equation} \label{ka1}
\kappa_1=1+\gamma.
\end{equation}

In this paper, we are exclusively concerned with light rays propagating in regions of spacetime  
such that $r\gg m$. The metric (\ref{bds}) can therefore be considered as a small perturbation 
of the background metric defined by  
\begin{equation} \label{opbck}
\gamma_{00}=1, \quad \gamma_{0i}=0, \quad \gamma_{ij}=-\left(1+\frac{2\kappa_1m}{r}\right)\delta_{ij},
\end{equation} 
the perturbation potentials being the first-order terms $\bar{h}^a_{\mu i}$ involved in the expansion of $\bar{h}_{\mu i}$ and the quantities $\bar{h}^{\kappa_n}_{\mu \nu}$ given by
\begin{equation} \label{k2h}
\bar{h}^{\kappa_n}_{00}=\bar{h}^{\kappa_n}_{0i}=0,\quad\bar{h}^{\kappa_n}_{ij}=
-\frac{2\kappa_n m^n}{r^n}\delta_{ij}
\end{equation} 
for $n\geq 2$.

It is easily inferred from Eqs. (\ref{kmn}) that the $\bar{k}^{\mu\nu}_{a}$ corresponding to the $\bar{h}^{a}_{\mu\nu}$ are given by
\begin{subequations} \label{ka}
\begin{align}
\bar{k}_{a}^{00}&=-\bar{h}_{00}^{a}=0, \label{ka00} \\
\bar{k}_{a}^{0i}&=\frac{r}{r+2\kappa_1m}\bar{h}_{0i}^{a}, \label{ka0i} \\
\bar{k}_{a}^{ij}&=-\frac{r^2}{(r+2\kappa_1m)^2}\bar{h}_{ij}^{a} \label{kaij}.
\end{align}
\end{subequations}

In the same way, the $\bar{k}^{\mu\nu}_{\kappa_n}$ corresponding to the $\bar{h}^{\kappa_n}_{\mu\nu}$ are
\begin{equation} \label{k2k}
\bar{k}^{00}_{\kappa_n}=\bar{k}^{0i}_{\kappa_n}=0,\quad \bar{k}_{\kappa_n}^{ij}
=\frac{2\kappa_nm^n}{r^{n-2}(r+2\kappa_1m)^2}\delta^{ij}.
\end{equation}

It follows from Eqs. (\ref{k2h}) that each parameter $\kappa_n$ generates a perturbation of 
order $m^n$. This property implies that only the perturbations due to $\kappa_1,\kappa_2$ and 
$\kappa_3$ can be consistently treated in the present work since the terms of order $m^4$ which 
are proportional to $(\kappa_2)^2$ are not taken into account by our procedure.  
 
The TTFs corresponding to the background metric (\ref{opbck}) are obtained in an exact closed 
form in \cite{Linet:2016}. We are led to distinguish three kinds of configurations according to 
the value of the angle $\psi_{AB}$ between the vectors $\bm{x}_A$ and $\bm{x}_B$.  
In what follows, $\psi_{AB}$ is unambiguously determined by the relations
\begin{equation} \label{psiAB}
\cos\psi_{AB}=\bm{n}_A.\bm{n}_B,\qquad 0\leq\psi_{AB}\leq\pi,
\end{equation}
where $\bm{n}_A$ and $\bm{n}_B$ are defined as
\begin{equation} \label{nanb}
\bm{n}_A=\frac{\bm{x}_A}{r_A},\qquad \bm{n}_B=\frac{\bm{x}_B}{r_B}.
\end{equation}

%%%%%%%%%%%%%%%%%%%%%%%%%%%%%%%%%%%%%%%%%%%%%%%%%%%%%%%%%
\subsection{Case where $\psi_{AB}=0$} \label{ssc:coll}
%%%%%%%%%%%%%%%%%%%%%%%%%%%%%%%%%%%%%%%%%%%%%%%%%%%%%%%%%

Since $\bm{n}_B=\bm{n}_A$ in this case, there exists one and only one zeroth-order light ray passing through $\bm{x}_A$ and $\bm{x}_B$, namely the radial null geodesic of the metric (\ref{opbck}) tangent to $\bm{n}_A$. We have $x^0_B-x^0_A=
\vert\int_{r_A}^{r_B}\sqrt{1+2\kappa_1m/r}\,dr\vert$ along this path. Hence the 
expression of the corresponding zeroth-order radial TTF: 
\begin{widetext}
\begin{equation} \label{TTr}
cT^{[rad]}(r_A\bm{n}_A,r_B\bm{n}_A)=\left\vert\sqrt{r_B(r_B+2\kappa_1m)}
-\sqrt{r_A(r_A+2\kappa_1m)}\right\vert+2\kappa_1m\left\vert\ln\left(\frac{\sqrt{r_B}
+\sqrt{r_B+2\kappa_1m}}{\sqrt{r_A}
+\sqrt{r_A+2\kappa_1m}}\right)\right\vert,
\end{equation} 
\end{widetext}
from which it is inferred that
\begin{equation} \label{grTr}
\left[\frac{c\partial T^{[rad]}}{\partial x^i}\right]_{\bm{x}=r\bm{n}_A}=\text{sgn}(r_B-r_A)\sqrt{1+\frac{2\kappa_1m}{r}}n^i_A.
\end{equation}

Denote by $c\Delta\mathcal{T}_{a}^{[rad]}$ the perturbation 
due to $\epsilon_a$ and by $c\Delta\mathcal{T}^{[rad]}_{\kappa_n}$ the  
perturbation due to $\kappa_n$ for $n\geq2$. Substituting for $\bar{k}_a^{\mu\nu}$ from Eqs. (\ref{ka}) into Eq. (\ref{dI1}), using Eq. (\ref{grTr}) and noting that Eq. (\ref{dl}) reads 
\begin{equation} \label{del1}
d\ell=\mbox{sgn}(r_B-r_A)\sqrt{1+2\kappa_1m/r}dr
\end{equation} 
when $r$ is the choosen parameter, it may be seen that Eq. (\ref{T1a}) takes the form
\begin{widetext}
\begin{equation} \label{DTr}
c\Delta\mathcal{T}^{[rad]}_{a}(r_A\bm{n}_A,r_B\bm{n}_A)
=-\int_{r_A}^{r_B}\left[\bar{h}^a_{i0}(r\bm{n}_A)n^i_A+\frac{1}{2}\text{sgn}(r_B-r_A)\sqrt{\frac{r}{r+2\kappa_1m}}\bar{h}_{ij}^a(r\bm{n}_A)n^i_An^j_A\right] dr. 
\end{equation}
\end{widetext}

Substituting $\bar{h}^{\kappa_n}_{i\alpha}$ for $\bar{h}^{a}_{i\alpha}$ into Eq. (\ref{DTr}) 
and taking Eqs. (\ref{k2h}) into account yields for $n\geq2$
\begin{equation} \label{Tknr}
c\Delta\mathcal{T}^{[rad]}_{\kappa_n}(r_A\bm{n}_A,r_B\bm{n}_A)=\kappa_nm^n\left\vert 
I_{n-1}(r_A,r_B)\right\vert,
\end{equation}
\newpage
where $I_n(r_A,r_B)$ is given by
\begin{equation} \label{Inab}
I_n(r_A,r_B)=\int_{r_A}^{r_B}\frac{dr}{r^n\sqrt{r(r+2\kappa_1m)}}.
\end{equation}

\begin{widetext}
We have for $n=2$ and $n=3$:
\begin{eqnarray} 
c\Delta\mathcal{T}^{[rad]}_{\kappa_2}(r_A\bm{n}_A,r_B\bm{n}_A)&=&
\frac{2\kappa_2 m^2 \vert r_B-r_A\vert}{r_A\sqrt{r_B(r_B+2\kappa_1m)}+r_B\sqrt{r_A(r_A+2\kappa_1m)}},
\label{DTk2r} \\
c\Delta\mathcal{T}^{[rad]}_{\kappa_3}(r_A\bm{n}_A,r_B\bm{n}_A)
&=&\frac{\kappa_3m^3\vert r_B^2-r_A^2\vert}{r_A^2(r_B-\kappa_1m)
\sqrt{r_B(r_B+2\kappa_1m)}+r_B^2(r_A-\kappa_1m)
\sqrt{r_A(r_A+2\kappa_1m)}}\nonumber \\
&&\times\left[1-\frac{2\kappa_1m}{3(r_A+r_B)}\left(1+\frac{r_A}{r_B}
+\frac{r_B}{r_A}\right)\right].
\label{DTk3r}
\end{eqnarray}
\end{widetext}

These exact expressions will be useful to prove the continuity of the perturbation terms proportional to $\kappa_2$ and to $\kappa_3$ (see the next subsection). In practice, however, 
owing to the fact that the quantities $I_n(r_A,r_B)$ can be expanded in series in powers 
of $m/r_A$ and $m/r_B$ which are convergent when $\text{inf}(r_A,r_B)>2\kappa_1m$,
we may content ourselves with approximate expressions as follow: 
\begin{widetext}
\begin{equation} \label{Tk2rd}
c\Delta\mathcal{T}^{[rad]}_{\kappa_2}(r_A\bm{n}_A,r_B\bm{n}_A)=\frac{\kappa_2m^2\vert r_B-r_A\vert}{r_Ar_B}\left[1-
\frac{\kappa_1m(r_A+r_B)}{2r_Ar_B}+O\left(\frac{m^2}{r_{AB}^2}\right)\right]
\end{equation}
and
\begin{equation} \label{Tk3rd}
c\Delta\mathcal{T}^{[rad]}_{\kappa_3}(r_A\bm{n}_A,r_B\bm{n}_A)
=\frac{\kappa_3m^3\vert r_B^2-r_A^2\vert}{2r_A^2r_B^2}\left[1+O\left(\frac{m}{r_{AB}}\right)\right],
\end{equation}
\end{widetext}
where 
\begin{equation} \label{rsAB}
r_{AB}=\inf(r_A,r_B).
\end{equation} 

%%%%%%%%%%%%%%%%%%%%%%%%%%%%%%%%%%%%%%%%%%%%%%%%%%%%%%%%%%%%%
\subsection{Case where $0<\psi_{AB}<\pi$} \label{Sssndca1}
%%%%%%%%%%%%%%%%%%%%%%%%%%%%%%%%%%%%%%%%%%%%%%%%%%%%%%%%%%%%%

Denote by $\bm{k}_{AB}$ the unit vector defined as 
\begin{equation} \label{defk}
\bm{k}_{AB}=\frac{\bm{n}_A\times\bm{n}_B}{\sin\psi_{AB}}. 
\end{equation}

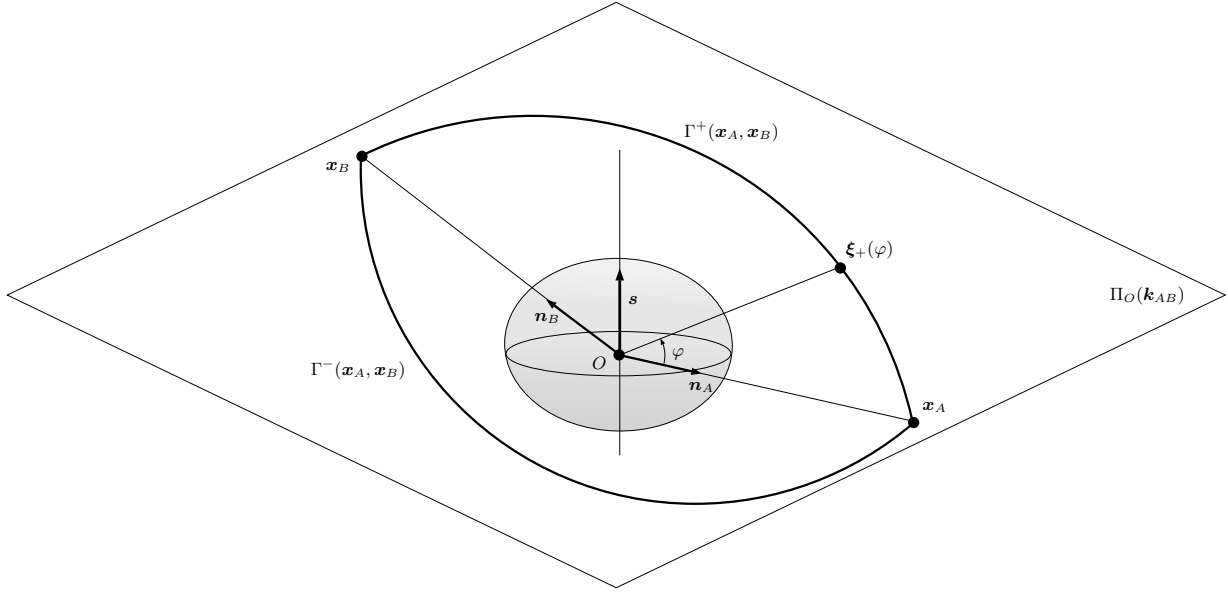
\begin{figure*}
\psscalebox{0.75 0.75}  %Change this value to rescale the drawing.
{
\begin{pspicture}(0,-8.34)(22.550276,2.06)
\definecolor{colour0}{rgb}{0.8,0.8,0.8}
\psrotate(10.850454, -4.020682){-1.4169099}{\psellipse[linecolor=black, linewidth=0.014, fillstyle=gradient, gradlines=2000, gradmidpoint=1, gradbegin=white, gradend=colour0, dimen=outer](10.850454,-4.020682)(2.025,1.5375757)}
\psellipse[linecolor=black, linewidth=0.014](10.850454,-4.180)(2.000,0.400)
\psdiamond[linecolor=black, linewidth=0.014, dimen=outer](10.811061,-3.14)(10.811061,5.2)
\psarc[linecolor=black, linewidth=0.04, dimen=outer](12.212467,-0.9128457){5.9300427}{177.92938}{310.77563}
\psrotate(9.33553, -6.8449244){39.466568}{\psarc[linecolor=black, linewidth=0.04, dimen=outer](9.33553,-6.8449244){6.8836365}{-26.98}{76.88401}}
\psdots[linecolor=black, dotsize=0.2](6.302121,-0.68)
\psdots[linecolor=black, dotsize=0.2](16.08212,-5.4)
\psline[linecolor=black, linewidth=0.014](10.862122,-4.200)(6.302121,-0.68)
\psline[linecolor=black, linewidth=0.014](10.857879,-4.1987877)(16.08212,-5.38)
\psdots[linecolor=black, dotsize=0.2](10.862122,-4.2)
\psdots[linecolor=black, dotsize=0.2](14.782122,-2.66)
\psline[linecolor=black, linewidth=0.014](10.882121,-4.2)(14.789697,-2.6358333)
\psline[linecolor=black, linewidth=0.04, arrowsize=0.05291667cm 2.0,arrowlength=1.4,arrowinset=0.0]{->}(10.862122,-4.190)(9.566904,-3.218)
\psline[linecolor=black, linewidth=0.04, arrowsize=0.05291667cm 2.0,arrowlength=1.4,arrowinset=0.0]{->}(10.877089,-4.2074337)(12.327154,-4.532566)
\psline[linecolor=black, linewidth=0.014](10.87319,-5.9820642)(10.87319,-0.56414795)(10.87319,-0.56414795)
\psline[linecolor=black, linewidth=0.05, arrowsize=0.05291667cm 2.0,arrowlength=1.4,arrowinset=0.0]{->}(10.87319,-4.15302)(10.87319,-2.66698)
\rput[bl](16.24212,-5.22){$\bm{x}_A$}
\rput[bl](5.6621214,-0.96){$\bm{x}_B$}
\rput[bl](12.100,-4.86){$\bm{n}_A$}
\rput[bl](9.362122,-3.64){$\bm{n}_B$}
\rput[bl](11.022121,-3.3){$\bm{s}$}
\psarc[linecolor=black, linewidth=0.014, dimen=outer, arrowsize=0.05291667cm 2.0,arrowlength=1.4,arrowinset=0.0]{->}(11.073155,-4.2085347){0.60048277}{-14.161057}{-330.0}
\rput[bl](11.800,-4.2773275){$\varphi$}
\rput[bl](5.3959146,-4.6269827){$\Gamma^{-}(\bm{x}_A,\bm{x}_B)$}
\rput[bl](12.015914,-0.40698275){$\Gamma^{+}(\bm{x}_A,\bm{x}_B)$}
\rput[bl](10.384811,-4.450){$O$}
\rput[bl](14.880,-2.4828448){$\bm{\xi}_{+}(\varphi)$}
\rput[bl](19.56091,-3.3052273){$\Pi_O(\bm{k}_{AB})$}
\end{pspicture}
}
\caption{\label{fig1} Representation of the zeroth-order light rays joining two given points in the generic case where the vector positions $\bm{x}_A$ and $\bm{x}_B$ are not collinear; these paths are confined in the plane $\Pi_O(\bm{k}_{AB})$ passing through the origin $O$ and containing $\bm{x}_A$ and $\bm{x}_B$.}
\end{figure*}

Since the background metric (\ref{opbck}) is spherically symmetric, its null geodesics paths joining $\bm{x}_A$ and $\bm{x}_B$ are confined in the plane $\Pi_O(\bm{k}_{AB})$ passing through the origin $O$ of the spatial coordinates and orthogonal to $\bm{k}_{AB}$. 
Denote by $\varphi$ the angle between $\bm{n}_A$ and the position vector $\bm{x}$ in the 
plane $\Pi_O(\bm{k}_{AB})$. Then, the first integrals of a null geodesic of the metric (\ref{opbck}) contained in $\Pi_O(\bm{k}_{AB})$ may be written as
\begin{eqnarray}
&&\frac{dx^0}{d\lambda}=E, \label{E}  \\
&&r(r+2\kappa_1m)\frac{d\varphi}{d\lambda}=J, \label{J} \\
&&\left(1+\frac{2\kappa_1m}{r}\right)
\left[\left(\frac{dr}{d\lambda}\right)^2+r^2\left(\frac{d\varphi}{d\lambda}\right)^2\right]
=E^2, \label{r}
\end{eqnarray}
where $E$ and $J$ are constants of the motion, $\lambda$ being an affine parameter.
 
The absolute value of the ratio $b$ defined by
\begin{equation} \label{b}
b=\frac{J}{E}
\end{equation}
is the intrinsic impact parameter of the corresponding zeroth-order light ray \cite{Chandrasekhar:1983}. Equation (\ref{r}) implies that the element $dl$ defined by 
Eq. (\ref{dl}) is related to the affine parameter $\lambda$ by the equation $dl=E d\lambda$ 
($\lambda$ may be chosen so that $E>0$). Hence
\begin{equation} \label{l}
dl=\frac{J}{b}d\lambda.
\end{equation}

Eliminating $J$ between Eq. (\ref{J}) and Eq. (\ref{l}) yields
\begin{equation} \label{dlph}
dl=\frac{r(r+2\kappa_1m)}{b} d\varphi.
\end{equation}

It is shown in \cite{Woude:1922} that there exist two and only two solutions to Eqs. (\ref{E})-(\ref{r}) passing through $\bm{x}_A$ and $\bm{x}_B$ when 
$0<\psi_{AB}<\pi$. These solutions, denoted by $\Gamma^{+}(\bm{x}_A,\bm{x}_B)$ and 
$\Gamma^{-}(\bm{x}_A,\bm{x}_B)$, are Keplerian hyperbolas with a focus coinciding with 
the origin $O$. By convention, $\Gamma^{+}(\bm{x}_A,\bm{x}_B)$ is the path along which $\varphi$ takes the values $0$ to $\psi_{AB}$, whereas $\Gamma^{-}(\bm{x}_A,\bm{x}_B)$ 
is the path along which $\varphi$ takes the values 0 to $\psi_{AB}-2\pi$. For the sake of brevity, we put
\begin{equation} \label{psipm}
\psi_{AB}^{+}=\psi_{AB},\qquad \psi_{AB}^{-}=\psi_{AB}-2\pi.
\end{equation}

The TTFs corresponding to $\Gamma^{+}(\bm{x}_A,\bm{x}_B)$ and to 
$\Gamma^{-}(\bm{x}_A,\bm{x}_B)$ will be denoted by $T^{+}(\bm{x}_A,\bm{x}_B)$ 
and $T^{-}(\bm{x}_A,\bm{x}_B)$, respectively. It will be relevant in some calculations 
to note that these functions satisfy the eikonal equation 
\begin{equation} \label{eikss}
\delta^{ij}\frac{c\partial T^{\pm}(\bm{x}_A,\bm{x})}{\partial x^i}
\frac{c\partial T^{\pm}(\bm{x}_A,\bm{x})}{\partial x^j}=1+\frac{2\kappa_1m}{r},
\end{equation}
which is straiforwardly inferred from Eq. (\ref{eikT0}) when Eqs. (\ref{opbck}) are taken into account.

Let us denote the parametric equations of $\Gamma^{+}(\bm{x}_A,\bm{x}_B)$ and 
$\Gamma^{-}(\bm{x}_A,\bm{x}_B)$ as functions of $\varphi$ by $\bm{\xi}_{+}(\varphi)$ 
and $\bm{\xi}_{-}(\varphi)$, respectively. Substituting for $\bar{k}_a^{\mu\nu}$ from Eqs. (\ref{ka}) in Eq. (\ref{dI1}), using Eq. (\ref{T1a}) and taking Eq. (\ref{dlph}) into account yields an expression as follows for $c\Delta\mathcal{T}_{a}^{\pm}(\bm{x}_A,\bm{x}_B)$:
\begin{widetext}
\begin{equation} \label{DTpm}
c\Delta\mathcal{T}_{a}^{\pm}(\bm{x}_A, \bm{x}_B)=-\frac{1}{b_{\pm}}\int_{0}^{\psi_{AB}^{\pm}}
\!r^2\left[\bar{h}^{a}_{0i}(\bm{x})\frac{c\partial T^{\,\pm}(\bm{x}_A,\bm{x})}{\partial x^i}
+\frac{r\bar{h}^{a}_{ij}(\bm{x})}{2(r+2\kappa_1m)}\frac{c\partial T^{\,\pm}(\bm{x}_A,\bm{x})}{\partial x^i}\frac{c\partial T^{\,\pm}(\bm{x}_A,\bm{x})}{\partial x^j}\right]_{\bm{x}=\bm{\xi}_{\pm}(\varphi)}\!d\varphi, 
\end{equation}
where $b_{+}$ (resp. $b_{-}$) is the algebraic value of the impact parameter of $\Gamma^{+}(\bm{x}_A,\bm{x}_B)$ (resp. $\Gamma^{-}(\bm{x}_A,\bm{x}_B)$). We have (see Eq. 
(31) in \cite{Linet:2016}) 
\begin{equation} \label{bpm}
b_{\pm}=\frac{r_Ar_B\displaystyle\sqrt{1-\cos\psi_{AB}}}{2R_{AB}}\left\lbrack
\sqrt{1+\cos \psi_{AB}+\frac{2\kappa_1m(r_A+r_B-R_{AB})}{r_Ar_B}}\pm\sqrt{1+\cos \psi_{AB}+\frac{2\kappa_1m(r_A+r_B+R_{AB})}{r_Ar_B}}\right\rbrack,   
\end{equation}
\end{widetext}
where $R_{AB}$ is defined as
\begin{equation} \label{RAB}
R_{AB}=\vert\bm{x}_B-\bm{x}_A\vert=\sqrt{r_A^2+r_B^2-2r_Ar_B
\cos\psi_{AB}}. 
\end{equation} 

Note that $b_{+}>0$ and $b_{-}<0$. 

The zeroth-order time transfer functions $T^{+}$ and $T^{-}$ corresponding to the background 
metric (\ref{opbck}) are given by Eq. (39) in \cite{Linet:2016}: 
\begin{widetext}
\begin{eqnarray}
cT^{\,\pm}(\bm{x}_A,\bm{x}_B)=&&\frac{1}{2}\left(\sqrt{r_A+r_B+R_{AB}}\sqrt{r_A+r_B+R_{AB}+4\kappa_1m}
\mp\sqrt{r_A+r_B-R_{AB}}\sqrt{r_A+r_B-R_{AB}+4\kappa_1m} \right)
  \nonumber \\
&&\mbox{}+2\kappa_1m\ln\left(\frac{\displaystyle\sqrt{r_A+r_B+R_{AB}+4\kappa_1m}
+\sqrt{r_A+r_B+R_{AB}}}
{\displaystyle\sqrt{r_A+r_B-R_{AB}+4\kappa_1m}\pm \sqrt{r_A+r_B-R_{AB}}}\right).\quad  \label{T0pm}
\end{eqnarray}

Substituting $\bm{x}$ for $\bm{x}_B$ into Eq. (\ref{T0pm}) and differentiating with respect 
to $x^i$ yields 
\begin{equation} 
\frac{c\partial T^{\pm}(\bm{x}_A,\bm{x})}{\partial x^i}=\frac{1}{2}\left\{\left[X^{+}(\bm{x}_A,\bm{x})\pm X^{-}(\bm{x}_A,\bm{x})\right]\frac{x^i-x_A^i}{\vert\bm{x}-\bm{x}_A\vert}+\left[X^{+}(\bm{x}_A,\bm{x})\mp X^{-}(\bm{x}_A,\bm{x})\right]\frac{x^i}{r}\right\},
\label{grTb}
\end{equation}
\end{widetext}
where $X^{+}(\bm{x}_A,\bm{x})$ and $X^{-}(\bm{x}_A,\bm{x})$ are given by
\begin{equation} \label{Xpm}
X^{\pm}(\bm{x}_A,\bm{x})=\sqrt{1+\frac{4\kappa_1m}{r_A+r\pm\vert\bm{x}
-\bm{x}_A\vert}}.
\end{equation}

The parametric functions $\bm{\xi}_{\pm}(\varphi)$ describing the hyperbolas $\Gamma^{+}(\bm{x}_A,\bm{x}_B)$ and $\Gamma^{-}(\bm{x}_A,\bm{x}_B)$ can be written in the 
form
\begin{equation} \label{x0phi}
\bm{\xi}_{\pm}(\varphi)=r_{\pm}(\varphi)[\cos\varphi\,\bm{n}_A+\sin\varphi\,(\bm{k}_{AB}\times\bm{n}_A)],
\end{equation}
where
\begin{equation} \label{rpm}
r_{\pm}(\varphi)=\frac{p_{\pm}}{1+e_{\pm}\cos (\varphi-\varphi^{\pm}_P)},
\end{equation}
the parameters $p_{\pm}$ and the eccentricities $e_{\pm}$ being given by relations as follow 
(see \cite{Linet:2016}):
\begin{equation} \label{ppm}
p_{\pm}=\frac{b^2_{\pm}}{\kappa_1 m},\quad\quad e_{\pm}=\frac{\vert b_{\pm}\vert}{\kappa_1 m}\sqrt{1+\frac{\kappa_1^2 m^2}{b^2_{\pm}}},
\end{equation}
and $\varphi^{+}_P$ (resp. $\varphi^{-}_P$) denoting the value of $\varphi$ at the pericenter of $\Gamma^{+}(\bm{x}_A,\bm{x}_B)$ (resp. $\Gamma^{-}(\bm{x}_A,\bm{x}_B)$). 
Given that $\varphi^{+}_P$ and $\varphi^{-}_P$ are determined by the equations which express that $\Gamma^{+}(\bm{x}_A,\bm{x}_B)$ and $\Gamma^{-}(\bm{x}_A,\bm{x}_B)$ pass through $\bm{x}_A$ and $\bm{x}_B$, namely
\begin{subequations}
\label{rarb}
\begin{eqnarray} 
&&\frac{p_{\pm}}{e_{\pm}r_A}-\frac{1}{e_{\pm}}=\cos\varphi^{\pm}_P, \label{rarba}\\
&&\frac{p_{\pm}}{e_{\pm}r_B}-\frac{1}{e_{\pm}}=\cos(\psi_{AB}-\varphi^{\pm}_P),  \label{rarbb}
\end{eqnarray}
\end{subequations}
it is easily seen that $1/r_{\pm}(\varphi)$ can be written as
\begin{widetext}
\begin{equation} \label{traj}
\frac{1}{r_{\pm}(\varphi)}=\frac{\kappa_1m}{b^2_{\pm}}+\frac{1}{\sin\psi_{AB}}
\left[\left(\frac{1}{r_A}-\frac{\kappa_1m}{b^2_{\pm}}\right)\sin(\psi_{AB}-\varphi)
+\left(\frac{1}{r_B}-\frac{\kappa_1m}{b^2_{\pm}}\right)\sin\varphi\right].
\end{equation}
\end{widetext}

Equations (\ref{bpm}), (\ref{grTb})-(\ref{Xpm}), (\ref{ppm}) and (\ref{traj}) constitute the ingredients allowing to write the explicit expression of the integral appearing in the right hand 
side of Eq. (\ref{DTpm}) when the perturbation terms $\bar{h}^{a}_{\mu\nu}$ are known. The computation of $c\Delta\mathcal{T}^{\pm}_{a}$ will be greatly simplified if $\bar{h}^a_{ij}=\bar{h}^a\delta_{ij}$, since it results from Eq. (\ref{eikss}) that
\begin{equation} \label{hkss}
\bar{h}^a_{ij}(\bm{x})\frac{c\partial T^{\pm}}{\partial x^i}
\frac{c\partial T^{\pm}}{\partial x^j}=\bar{h}^a(\bm{x})\left(1+\frac{2\kappa_1m}{r}\right).
\end{equation}
In particular, substituting for 
$\bar{h}_{\mu\nu}^{\kappa_n}$ from Eqs. (\ref{k2h}) into Eq. (\ref{DTpm}) straightforwardly leads to
\begin{equation} \label{Tka2}
c\Delta\mathcal{T}^{\pm}_{\kappa_2}(\bm{x}_A,\bm{x}_B)=\frac{\kappa_2m^2}{b_{\pm}}
\psi_{AB}^{\pm}
\end{equation}
and
\begin{eqnarray} 
\!\!\!c\Delta\mathcal{T}^{\pm}_{\kappa_3}(\bm{x}_A,\bm{x}_B)&=&\frac{\kappa_3m^3}{b_{\pm}^2}
\bigg\lbrack\frac{\sin\psi_{AB}^{\pm}}{1+\cos\psi_{AB}^{\pm}}\left(\frac{b_{\pm}}{r_A}
+\frac{b_{\pm}}{r_B}\right)  \nonumber \\
&&+\frac{\kappa_1m}{b_{\pm}}\left(\psi_{AB}^{\pm}-\frac{2\sin\psi_{AB}^{\pm}}{1+\cos\psi_{AB}^{\pm}}\right)\bigg\rbrack.\nonumber \\
&& \label{Tka3}
\end{eqnarray} 

%%%%%%%%%%%%%%%%%%%%%%%%%%%%%%%%%%%%%%%%%%%%%%%%%%%%%%%%%%%%%%%%%%%%%%%%%%
\subsection{Case where $\psi_{AB}=\pi$} \label{Samb}
%%%%%%%%%%%%%%%%%%%%%%%%%%%%%%%%%%%%%%%%%%%%%%%%%%%%%%%%%%%%%%%%%%%%%%%%%%
 
Let $\bm{k}$ be any unit vector orthogonal to the common direction of $\bm{n}_A$ and $\bm{n}_B$. The plane $\Pi_O(\bm{k})$ passing through $O$ and orthogonal to $\bm{k}$ 
contains two zeroth-order null geodesics joining $\bm{x}_A$ and $\bm{x}_B$. The arbitrariness in 
the choice of $\Pi_O(\bm{k})$ implies that our procedure leads in this case to a double 
infinity of TTFs whose perturbed part depends on $\bm{k}$. Explicit examples of this dependence  are given in Subsect. \ref{SABo}. The zeroth-order terms $cT^{\pm}$ are simply obtained by 
setting $R_{AB}=r_A+r_B$ in Eq. (\ref{T0pm}), which gives
\begin{eqnarray}
cT^{\pm}(r_A\bm{n}_A,-r_B\bm{n}_A)
&=&(r_A+r_B)\sqrt{1+\frac{2\kappa_1m}{r_A+r_B}} \nonumber \\
&&-\kappa_1m\ln\frac{2\kappa_1m}{r_A+r_B}\nonumber \\
&&+2\kappa_1m\ln\left(1+\sqrt{1+\frac{2\kappa_1m}{r_A+r_B}}\right)\nonumber \\
&& \label{T0pi}
\end{eqnarray}

Equation (\ref{bpm}) reduces to 
\begin{equation} \label{bpmpi}
b_{\pm}=\pm\sqrt{\frac{2\kappa_1mr_Ar_B}{r_A+r_B}}.
\end{equation}

Substituting for $b_{\pm}$ from Eq. (\ref{bpmpi}) into Eqs. (\ref{ppm}) yields
expressions as follow for $p_{\pm}$ and $e_{\pm}$:
\begin{eqnarray}
p_{\pm}&=&\frac{2r_Ar_B}{r_A+r_B},  \label{ppmpi}\\
e_{\pm}&=&\sqrt{\frac{2r_Ar_B}{\kappa_1m(r_A+r_B)}+1}.\label{epmpi}
\end{eqnarray}
Using Eq. (\ref{ppmpi}), it is easily deduced from Eqs. (\ref{rarb}) that
\begin{equation} \label{perpi}
e_{\pm}\cos\varphi^{\pm}_P=\frac{r_B-r_A}{r_A+r_B}.
\end{equation}
This relation determines the angular position of the respective pericenters of $\Gamma^{+}(\bm{x}_A,\bm{x}_B;\bm{k})$ and $\Gamma^{-}(\bm{x}_A,\bm{x}_B;\bm{k})$ when the geometrically obvious conditions $0<\varphi^{+}_P<\pi$ and $\varphi^{-}_P=-\varphi^{+}_P$ are 
taken into account. Equations (\ref{epmpi}) and (\ref{perpi}) imply that
\begin{equation} \label{snpi}
e_{\pm}\sin\varphi^{\pm}_P=\pm\sqrt{\frac{2r_Ar_B}{\kappa_1m(r_A+r_B)}
\left(1+\frac{2\kappa_1m}{r_A+r_B}\right)}.
\end{equation}

To finish, Eqs. (\ref{rpm}), (\ref{ppmpi}), (\ref{perpi}) and (\ref{snpi}) lead to 
\begin{eqnarray} \label{trj1}
\!\!\!\!\frac{1}{r_{\pm}(\varphi)}&=&\frac{r_A+r_B}{2r_Ar_B}
+\frac{r_B-r_A}{2r_Ar_B}\cos\varphi\nonumber \\
&&\pm\sqrt{\frac{r_A+r_B}{2\kappa_1mr_Ar_B}
\left(1+\frac{2\kappa_1m}{r_A+r_B}\right)}\sin\varphi.
\end{eqnarray}

Equations (\ref{T0pm})-(\ref{Xpm}), (\ref{bpmpi}), (\ref{ppmpi}) and (\ref{trj1}) constitute the whole set of relations enabling one to get the explicit expressions of the 
integrals involved in Eq. (\ref{DTpm}) when the emitter and the receiver are located in diametrically opposite directions. The contributions due to $\kappa_2$ and $\kappa_3$ are given 
by formulas as follow:
\begin{eqnarray}
c\Delta\mathcal{T}^{\pm}_{\kappa_2}(\bm{x}_A,\bm{x}_B)&=&\frac{\kappa_2\pi m}{\sqrt{2\kappa_1}}
\sqrt{\frac{m(r_A+r_B)}{r_Ar_B}}, \label{Tk2l} \\ 
c\Delta\mathcal{T}^{\pm}_{\kappa_3}(\bm{x}_A,\bm{x}_B)&=&\frac{\kappa_3m^2}{\kappa_1}
\frac{r_A+r_B}{r_Ar_B}\Bigg\lbrack\sqrt{1+\frac{2\kappa_1m}{r_A+r_B}}\nonumber \\
&&+\frac{\pi}{2}
\sqrt{\frac{\kappa_1m(r_A+r_B)}{2r_Ar_B}}\Bigg\rbrack. \label{Tk3l}
\end{eqnarray}

For $\psi_{AB}$ close to $\pi$, Eq. (\ref{bpm}) implies
\begin{eqnarray}
b_{\pm}&=&\pm\sqrt{\frac{2\kappa_1mr_Ar_B}{r_A+r_B}}\nonumber \\
&&+\frac{r_Ar_B}{2(r_A+r_B)}\sqrt{1+\frac{2\kappa_1m}{r_A+r_B}}
(\pi-\psi_{AB})\nonumber \\
&&+O\left[(\pi-\psi_{AB})^2\right]. \label{lbpm}
\end{eqnarray}

Using Eq. (\ref{lbpm}), it may be seen that Eq. (\ref{Tk2l}) and Eq. (\ref{Tk3l}) are 
respectively the limit of Eq. (\ref{Tka2}) and Eq. (\ref{Tka3}) when $\psi_{AB}\rightarrow \pi$.
It is thus explicitly checked that the functions $c\Delta\mathcal{T}^{\pm}_{\kappa_2}$ and $c\Delta\mathcal{T}^{\pm}_{\kappa_3}$ calculated above in the different possible 
configurations are continuous whatever $\bm{x}_A$ and $\bm{x}_B$.

%%%%%%%%%%%%%%%%%%%%%%%%%%%%%%%%%%%%%%%%%%%%%%%%%%%%%%%%%%%%%%%%%%%%%%%%%%%%%%%%%%%%%%
\subsection{Existence of two regimes} \label{ssec:reg}
%%%%%%%%%%%%%%%%%%%%%%%%%%%%%%%%%%%%%%%%%%%%%%%%%%%%%%%%%%%%%%%%%%%%%%%%%%%%%%%%%%%%%%

Our method is valid only if the zeroth-order light ray along which the perturbation terms are calculated is entirely located in a region of weak field, which amounts to supposing that 
$r\gg m$ at any point of the path. Since it is assumed in this paper that 
$\text{inf}(r_A,r_B)\gg m$, the weak-field condition can be violated only if the null geodesic path joining $\bm{x}_A$ and $\bm{x}_B$ passes through a pericenter whose radial coordinate is close to $m$. Denote by $r_P^{+}$ (resp. $r_P^{-}$) the value of $r$ at 
the pericenter of $\Gamma^{+}(\bm{x}_A,\bm{x}_B)$ (resp. 
$\Gamma^{-}(\bm{x}_A,\bm{x}_B)$). It follows from Eqs. (\ref{rpm}) and (\ref{ppm}) that 
\begin{equation} \label{rPpm}
r_P^{\pm}=\sqrt{b_{\pm}^2+\kappa_1^2m^2}-\kappa_1m.
\end{equation}
Using Eqs. (\ref{bpm}) and (\ref{rPpm}), a discussion which can be summarized as follows leads to distinguish at least two regimes (see Ref. \cite{Linet:2016} for the details).

1. {\it First regime.}-- If $\psi_{AB}$ is such that \cite{arccos} 
\begin{equation} \label{reg1}
\arccos\left[\frac{\kappa_1m(r_A+r_B)}{r_Ar_B}-1\right]\lesssim\psi_{AB}\leq\pi,  
\end{equation}
then $\Gamma^{+}(\bm{x}_A,\bm{x}_B)$ and $\Gamma^{-}(\bm{x}_A,\bm{x}_B)$ are located in the region where the weak field approximation is valid. This regime is typically encountered in a configuration of gravitational lensing.

2. {\it Second regime.}-- If $\psi_{AB}$ is such that 
\begin{equation}  \label{reg2}
0<\psi_{AB}\ll\arccos\left[\frac{\kappa_1m(r_A+r_B)}{r_Ar_B}-1\right],
\end{equation} 
it may be seen that $r_P^{-}\thicksim \kappa_1m$, a value which is not compatible with the weak-field approximation. As a consequence, only the perturbation terms $c\Delta\mathcal{T}_{a}^{+}$ and $c\Delta\mathcal{T}_{\kappa_n}^{+}$ can be considered as legitimate in the context of our procedure when condition (\ref{reg2}) is met. 

The second regime is typically encountered when we make $\psi_{AB}$ tend to $0$. The computation of the limits of $c\Delta\mathcal{T}^{+}_{\kappa_2}$ and $c\Delta\mathcal{T}^{+}_{\kappa_3}$ in 
this case is easy if we note that
\begin{eqnarray} 
b_{+}&=&\frac{r_A\sqrt{r_B(r_B+2\kappa_1m)}
+r_B\sqrt{r_A(r_A+2\kappa_1m)}}{2\vert r_B-r_A\vert}\psi_{AB}\nonumber \\
&&\times[1+O(\psi_{AB}^2)] \label{lbp}
\end{eqnarray}
when $\psi_{AB}$ is close to 0. Substituting for $b_{+}$ from Eq. (\ref{lbp}) into Eq. (\ref{Tka2}), it appears that $\lim_{\psi_{AB}\rightarrow 0}c\Delta\mathcal{T}^{+}_{\kappa_2}(\bm{x}_A,\bm{x}_B)$ coincides with the right-hand side of Eq. (\ref{DTk2r}). The continuity of the perturbation due to $\kappa_2$ as $\psi_{AB}\rightarrow 0$ is thus directly proved. A similar conclusion can be drawn for the perturbation due to $\kappa_3$. Indeed, Eqs. (\ref{Tka3}) and (\ref{lbp}) lead to an expression of $\lim_{\psi_{AB}\rightarrow 0}c\Delta\mathcal{T}^{+}_{\kappa_3}(\bm{x}_A,\bm{x}_B)$ which can be checked with some calculations to be identical to the right-hand side of Eq. (\ref{DTk3r}).

%%%%%%%%%%%%%%%%%%%%%%%%%%%%%%%%%%%%%%%%%%%%%%%%%%%%%%%%%%%%%%%%%%%%%%%%%%%%%%%%
\subsection{TTFs in static, spherically symmetric spacetimes} \label{S:sss}
%%%%%%%%%%%%%%%%%%%%%%%%%%%%%%%%%%%%%%%%%%%%%%%%%%%%%%%%%%%%%%%%%%%%%%%%%%%%%%%%

When the source of the gravitational field is a non-rotating spherically symmetric body, the perturbation potentials $\bar{h}_{i\alpha}$ vanish. The TTFs provided by our procedure then reduce to the functions $\mathcal{T}^{\pm}_{sph}$ given by 
\begin{equation} \label{Tsppm}
c\mathcal{T}^{\pm}_{sph}=cT^{\pm}+c\Delta\mathcal{T}_{\kappa_{2}}^{\pm}+
c\Delta\mathcal{T}_{\kappa_{3}}^{\pm}.
\end{equation}

It is of interest to compare the expansion of $\mathcal{T}^{+}_{sph}$ in ascending powers of $m$ with the expression of the TTF obtained in \cite{Linet:2013}. It follows from Eq. (\ref{T0pm}) 
that $cT^{+}$ can be expanded into a convergent series in powers of $m$ if and only if 
\begin{equation} \label{cv1}
r_A+r_B-R_{AB}\geq 4\kappa_1m.
\end{equation}
Since we assume $\inf(r_A,r_B)\gg m$, it is easily seen that condition (\ref{cv1}) is equivalent to 
\begin{equation} \label{cv2}
0\leq\psi_{AB}\leq\psi_{m}(r_A,r_B),
\end{equation}
where $\psi_{m}(r_A,r_B)$ is defined as
\begin{equation} \label{psm}
\psi_{m}(r_A,r_B)=\arccos
\left[\frac{4\kappa_1m(r_A+r_B-2\kappa_1m)}{r_Ar_B}-1\right].
\end{equation}
On the other hand, it is straightforwardly inferred from Eq. (\ref{bpm}) that $b_{+}$ can be developed into a convergent series in powers of $m$ if and only if 
\begin{equation} \label{cv3}
\frac{2\kappa_1m(r_A+r_B+R_{AB})}{r_Ar_B(1+\cos\psi_{AB})}\leq1,
\end{equation}
a condition which can be checked as being satisfied when $\psi_{AB}\leq\psi_m(r_A,r_B)$. So, using the expansion of $cT^{+}$ in powers of $m$ given in \cite{Linet:2016} and substituting for $1/b_{+}$ from
\begin{eqnarray}
\frac{1}{b_{+}}&=&\frac{R_{AB}}{r_Ar_B\sin\psi_{AB}} \nonumber \\
&&\times\left[1-\frac{\kappa_1m}{1+\cos\psi_{AB}}\frac{r_A+r_B}{r_Ar_B}
+O\left(\frac{m^2}{r_{AB}^2}\right)\right]\label{ivbp}
\end{eqnarray}
into Eqs. (\ref{Tka2}) and (\ref{Tka3}), it is rigorously shown that $c\mathcal{T}_{sph}^{+}$ can 
be expanded as
\begin{equation} \label{Tspp}
c\mathcal{T}_{sph}^{+}(\bm{x}_A,\bm{x}_B)=c\mathcal{T}_{sph}^{\ast}(\bm{x}_A,\bm{x}_B)+O\left(\frac{m^4}{r_{AB}^4}\right)
\end{equation}
with 
\begin{eqnarray}
c\mathcal{T}_{sph}^{\ast}(\bm{x}_A,\bm{x}_B)&=&R_{AB}+\kappa_1m\ln\left(\frac{r_A+r_B
+R_{AB}}{r_A+r_B-R_{AB}}\right) \nonumber \\
&&+\frac{m^2R_{AB}}{r_Ar_B}\left(
\frac{\kappa_2\psi_{AB}}{\sin\psi_{AB}}-\frac{\kappa_1^2}{1+\cos\psi_{AB}}\right)
\nonumber\\
&&+\frac{m^3R_{AB}(r_A+r_B)}{r_A^2r_B^2(1+\cos\psi_{AB})} \nonumber \\
&&\;\;\;\times\left(\kappa_3-\frac{\kappa_1\kappa_2\psi_{AB}}{\sin\psi_{AB}}
+\frac{\kappa_1^3}{1+\cos\psi_{AB}}\right) \nonumber \\
&&\label{dTsp}
\end{eqnarray}
if and only if condition (\ref{cv2}) is satisfied. 

We thus recover Eqs. (17), (62), (86) and (87) obtained in \cite{Linet:2013}, while bringing a 
very important precision since we now specify the condition of convergence of the expansion provided by Eqs. (\ref{Tspp}) and (\ref{dTsp}).

As a final remark, it may be noted that the limit of the right-hand side of Eq. (\ref{dTsp}) when $\psi_{AB}\rightarrow0$ is the development of
$c\mathcal{T}_{sph}^{[rad]}(r_A\bm{n}_A,r_B\bm{n}_B)$ obtained by expanding $\left\vert\int_{r_A}^{r_B}\sqrt{\mathcal{U}(r)}dr\right\vert$ into a series of powers of $m$ up to and including the third order in $m$ (see Eqs. (19)-(22) in \cite{Linet:2013}). This conclusion is directly inferred from $\lim_{\psi_{AB}\rightarrow0}R_{AB}=\vert r_B-r_A\vert$.

%%%%%%%%%%%%%%%%%%%%%%%%%%%%%%%%%%%%%%%%%%%%%%%%%%%%%%%%%%%%%%%%%%%%%%%%%%%%%%%%
\section{Application to a spinning axisymmetric body} \label{sec:rot}
%%%%%%%%%%%%%%%%%%%%%%%%%%%%%%%%%%%%%%%%%%%%%%%%%%%%%%%%%%%%%%%%%%%%%%%%%%%%%%%%

The formulas obtained in Sec. \ref{sec:pSchw} can be applied when the gravitational 
field is generated by an isolated, axisymmetric body of mass $M$ slowly rotating about its axis 
of symmetry with a constant angular velocity $\bm{\omega}$. The problem is treated here within 
the post-Newtonian framework for fully conservative theories. Modeling the light propagation for the current experiments performed in the Solar System or the astrometry at a level of 1 microarcsecond requires to retain only the terms of order $c^{-2}$ and $c^{-3}$ in the metric \cite{Klioner:1992,Klioner:2003}. So we will content ourselves with a physical metric having the form
\begin{eqnarray}
ds^2&=&\left(1-\frac{2W}{c^2}\right) (dx^0)^2+\frac{4\kappa_1}{c^3}(\bm{W}.\bm{d}\bm{x})dx^0 \nonumber \\
&&\mbox{}-\left(1+\frac{2\gamma W}{c^2}\right)\delta_{ij}dx^idx^j,\label{ds2}
\end{eqnarray} 
where $W$ is the Newtonian-like potential of the body, namely 
\begin{equation} \label{W}
W(\bm{x})=G\int_{\mathcal{D}}\frac{\rho^{\ast}(\bm{x}')}{\vert\bm{x}-\bm{x}'\vert}d^3\bm{x}',
\end{equation}
and $\bm{W}$ is the gravitomagnetic vector defined as
\begin{equation} \label{hd}
\bm{W}(\bm{x})=G\int_{\mathcal{D}}\frac{\rho^{\ast}(\bm{x}')(\bm{\omega}\times\bm{x}')}{\vert\bm{x}-\bm{x}'\vert}d^3\bm{x}',
\end{equation}
$\rho^{\ast}$ being the conserved mass density of the body (see, e.g., \cite{Will:2018}) and $\mathcal{D}$ the volume occupied by the matter.

The origin $O$ of the spatial coordinates $x^i$ is assumed to be situated on the axis of symmetry. Then, outside any sphere of radius $r_0$ centered on $O$ and encompassing the central body, 
$W/c^2$ may be represented by the well-known multipole expansion
\begin{equation} \label{Wexp}
\frac{W(\bm{x})}{c^2}=\frac{m}{r}\left[1-\sum_{n=1}^{\infty}J_n\left(\frac{r_0}{r}\right)^n P_n(\bm{s}.\bm{n})\right],
\end{equation}
where $\bm{s}$ is a unit vector parallel to the axis of symmetry, $\bm{n}$ is defined as
\begin{equation} \label{n}
\bm{n}=\frac{\bm{x}}{r},
\end{equation}
$P_n$ denotes the Legendre polynomial of degree $n$ and $J_n$ is the corresponding mass multipole moment of the body. In most cases, $r_0$ can be chosen as the equatorial radius of the body. 
Of course, we have $J_1=0$ if the origin $O$ coincides with the center of mass.

Let $\bm{S}$ be the intrinsic angular momentum of the body rotating about its axis of symmetry. 
We have
\begin{equation} \label{anm}
\bm{S}=S\bm{s},
\end{equation} 
where $S=I\omega$, $I$ being the moment of inertia of the body about its axis of symmetry and 
$\omega$ being defined by $\bm{\omega}=\omega\bm{s}$. By analogy with the Kerr metric, it is convenient to introduce the quantity $a$ having the dimension of a length defined as 
\begin{equation} \label{a}
a=\frac{S}{Mc}.
\end{equation}
With this notation, the gravitomagnetic vector $\bm{W}/c^3$ may be written in the form (see, e.g.,  \cite{Linet:2002} and references therein)
\begin{equation} \label{exph}
\frac{\bm{W}(\bm{x})}{c^3}=\frac{ma(\bm{s}\times\bm{x})}{2r^3}\left[1-\sum_{n=1}^{\infty}K_n\left(\frac{r_{0}}{r}\right)^n\!P'_{n+1}(\bm{s}.\bm{n})\right],
\end{equation}
where $P'_{n+1}(x)$ denotes the derivative of the Legendre polynomial $P_{n+1}(x)$ with respect 
to $x$ and each $K_n$ is a constant depending on the matter distribution inside the body.

For light rays, the conformal metric obtained by dividing the right-hand side of Eq. (\ref{ds2}) 
by $1-2W/c^2$ can be used in place of the initial metric. Since the terms of order $c^{-4}$ are neglected, the conformal metric to be considered here reduces to  
\begin{eqnarray}
d\bar{s}^2&=&(dx^0)^2 +\frac{4\kappa_1}{c^3}\left({\bm W}.{\bm d}\bm{x}\right)dx^0 \nonumber \\ &&\mbox{}-\left(1+\frac{2\kappa_1 W}{c^2}\right){\bm d}\bm{x}^2. \label{ds2op}
\end{eqnarray}

In accordance with the methodology developed in Section \ref{sec:pSchw}, the right-hand side of 
Eq. (\ref{ds2op}) is regarded as a stationary perturbation of the background metric defined by 
Eq. (\ref{opbck}). The perturbation parameters $\epsilon_{a}$ are then identified with the dimensionless quantities $J_n$, $a/m$ and $K_n$. The non-zero perturbations of the background metric may be written as follows:
\begin{equation} \label{hijJn}
\bar{h}_{ij}^{J_n}(\bm{x})=2\kappa_1m J_n\frac{r_0^n}{r^{n+1}}P_n(\bm{s}. \bm{n})\delta_{ij}
\end{equation}
for the contribution due to the multipole $J_n$,
\begin{equation} \label{h0iS}
\bar{\bm{h}}^{S}(\bm{x})=\frac{\kappa_1ma (\bm{s}\times\bm{n})}{r^2}
\end{equation}
for the dominant contribution due to $\bm{S}$ 
and 
\begin{equation} \label{h0iKn}
\bar{\bm{h}}^{K_n}(\bm{x})=-\kappa_1ma K_n (\bm{s}\times\bm{n})\frac{r_0^n}{r^{n+2}}P'_{n+1}(\bm{s}.\bm{n})
\end{equation}
for the contribution of the spin multipole $K_n$, 
where we use the vector notation
\begin{equation} \label{voff}
\bar{\bm{h}}^a=(\bar{h}^a_{0i}).
\end{equation}

%%%%%%%%%%%%%%%%%%%%%%%%%%%%%%%%%%%%%%%%%%%%%%%%%%%%%%%%%%%%%%%%%%%%%%%%%%%%%%%%%%%%%%%
\subsection{Case where $\psi_{AB}=0$} \label{Snen}
%%%%%%%%%%%%%%%%%%%%%%%%%%%%%%%%%%%%%%%%%%%%%%%%%%%%%%%%%%%%%%%%%%%%%%%%%%%%%%%%%%%%%%%

Substituting $\bar{h}_{ij}^{J_n}$ for $\bar{h}^a_{ij}$ into Eq. (\ref{DTr}) gives
\begin{eqnarray} 
c\Delta\mathcal{T}_{J_n}^{[rad]}(r_A\bm{n}_A,r_B\bm{n}_A)&=&-\kappa_1 m J_n r_0^n
\vert I_n(r_A,r_B)\vert \nonumber \\
&&\times P_n(\bm{s}\cdot\bm{n}_A), 
\label{cTJn}
\end{eqnarray}
where the quantities $I_n(r_A,r_B)$ are defined by Eq. (\ref{Inab}).   

Equation (\ref{grTb}) implies that the gradient of $c T^{+}(\bm{x}_A,\bm{x})$ and $\bm{x}$ are  collinear when $\bm{x}=r\bm{n}_A$. As a consequence, it follows from Eqs. (\ref{DTr}), (\ref{h0iS}) and (\ref{h0iKn}) that 
\begin{equation} \label{TSKra}
c\Delta\mathcal{T}_{S}^{[rad]}(r_A\bm{n}_A,r_B\bm{n}_A)=0 
\end{equation}
and
\begin{equation} \label{TSKrb}
c\Delta\mathcal{T}_{K_n}^{[rad]}(r_A\bm{n}_A,r_B\bm{n}_A)=0.
\end{equation}

Taking into account the relation
\begin{equation} \label{Inap}
I_n(r_A,r_B)=\frac{1}{n}\left(\frac{1}{r_A^n}-\frac{1}{r_B^n}\right)
\left[1+O\left(\frac{m}{r_{AB}}\right)\right]
\end{equation}
and Eqs. (\ref{TSKra})-(\ref{TSKrb}), it may be seen that Eqs. (\ref{TTr}) and  (\ref{cTJn}) lead to an expansion as follows:
\begin{widetext}
\begin{equation} \label{cTJa}
c\mathcal{T}^{[rad]}(r_A\bm{n}_A,r_B\bm{n}_A)=\vert r_B-r_A\vert
+\kappa_1m\left\{\left\vert\ln\frac{r_B}{r_A}\right\vert-\sum_{n=1}^{\infty}\frac{1}{n}
J_n\left\vert\frac{r_0^n}{r_A^n}-\frac{r_0^n}{r_B^n}\right\vert P_n(\bm{s}.\bm{n}_A)\right\}
+O\left(\frac{m^2}{r_{AB}^2}\right),
\end{equation} 
\end{widetext}
where $r_{AB}$ is defined by Eq. (\ref{rsAB}).

According to the assumption made at the beginning of this section, we may content ourselves 
with the approximation provided by Eq. (\ref{cTJa}). However, exact expressions or full expansions are necessary for checking the continuity of the perturbation functions associated with the mass 
and spin multipoles. In the next subsection, explicit expressions are obtained for the perturbation terms due to $J_1$, $J_2$ and $\bm{S}$ when $\bm{n}_A$ and $\bm{n}_B$ are not collinear. So, it will be essential to know the exact expressions of $c\Delta\mathcal{T}_{J_1}^{[rad]}$ and $c\Delta\mathcal{T}_{J_2}^{[rad]}$ which are directly inferred from (\ref{cTJn}), namely:
\begin{widetext}
\begin{eqnarray}
c\Delta\mathcal{T}_{J_1}^{[rad]}(r_A\bm{n}_A,r_B\bm{n}_A)&=&-2\kappa_1mJ_1r_0
\frac{\vert r_B-r_A\vert (\bm{s}.\bm{n}_A)}{r_A\sqrt{r_B(r_B+2\kappa_1m)}
+r_B\sqrt{r_A(r_A+2\kappa_1m)}}, \label{cTJ1r} \\
& & \nonumber \\
c\Delta\mathcal{T}_{J_2}^{[rad]}(r_A\bm{n}_A,r_B\bm{n}_A)&=&-\frac{1}{2}\kappa_1mJ_2r_0^2\frac{\vert r_B^2-r_A^2\vert 
[3(\bm{s}.\bm{n}_A)^2-1]}{r_A^2(r_B-\kappa_1m)\sqrt{r_B(r_B+2\kappa_1m)}
+r_B^2(r_A-\kappa_1m)\sqrt{r_A(r_A+2\kappa_1m)}} \nonumber \\
&&\mbox{}\times\left[1-\frac{2\kappa_1m}{3(r_A+r_B)}
\left(1+\frac{r_A}{r_B}+\frac{r_B}{r_A}\right)\right]. \label{cTJ2r}
\end{eqnarray}
\end{widetext}

%%%%%%%%%%%%%%%%%%%%%%%%%%%%%%%%%%%%%%%%%%%%%%%%%%%%%%%%%%%%%%%
\subsection{Case where $0<\psi_{AB}<\pi$} \label{sec:nnn0}
%%%%%%%%%%%%%%%%%%%%%%%%%%%%%%%%%%%%%%%%%%%%%%%%%%%%%%%%%%%%%%%

Substituting  $\bar{h}^{J_n}_{ij}$ for $\bar{h}^{a}_{ij}$ into Eq. (\ref{DTpm}) and taking 
into account Eq. (\ref{hkss}) yields an expression as follows for 
$c\Delta\mathcal{T}^{\pm}_{J_n}$:
\begin{equation} \label{TJn}
c\Delta\mathcal{T}_{J_n}^{\pm}(\bm{x}_A,\bm{x}_B)=-\frac{\kappa_1m}{b_{\pm}}J_nr_0^n
\mathcal{J}^{\pm}_n(\bm{x}_A, \bm{x}_B),
\end{equation}
where 
\begin{equation} \label{IJn}
\mathcal{J}^{\pm}_n(\bm{x}_A, \bm{x}_B)
=\int_{0}^{\psi_{AB}^{\pm}}P_n\left[\frac{\bm{s}\cdot\bm{\xi}_{\pm}(\varphi)}{r_{\pm}(\varphi)}\right]\frac{d\varphi}{r_{\pm}^{n-1}(\varphi)},
\end{equation}
$\bm{\xi}_{\pm}(\varphi)/r_{\pm}(\varphi)$ being directly inferred from Eq. (\ref{x0phi}) and 
$1/r_{\pm}(\varphi)$ being given by Eq. (\ref{traj}).

The contributions due to the gravitomagnetic potentials are given by 
\begin{equation} \label{iTS}
c\Delta\mathcal{T}_{S}^{\pm}(\bm{x}_A,\bm{x}_B)=-\kappa_1ma(\bm{s}.\bm{k}_{AB})\int_{0}^{\psi^{\pm}_{AB}}\frac{d\varphi}{r_{\pm}(\varphi)}
\end{equation}
and
\begin{equation} \label{iTKn}
c\Delta\mathcal{T}_{K_n}^{\pm}(\bm{x}_A,\bm{x}_B)=\kappa_1ma(\bm{s}.\bm{k}_{AB})K_n r_0^n
\mathcal{K}_n^{\pm}(\bm{x}_A,\bm{x}_B),
\end{equation}
where 
\begin{equation} \label{IKn}
\mathcal{K}^{\pm}_n(\bm{x}_A,\bm{x}_B)=\int_{0}^{\psi^{\pm}_{AB}}P'_{n+1}
\left[\frac{\bm{s}\cdot\bm{\xi}_{\pm}(\varphi)}{r_{\pm}(\varphi)}\right]\frac{d\varphi}{r_{\pm}^{n+1}(\varphi)}.
\end{equation} 

The integrals involved in Eqs. (\ref{IJn}), (\ref{iTS}) and (\ref{IKn}) can be calculated with 
any symbolic computer program. In this work, we only calculate $c\Delta\mathcal{T}_{J_1}^{\pm}$, $c\Delta\mathcal{T}_{J_2}^{\pm}$ and $c\Delta\mathcal{T}_{S}^{\pm}$. 

Using Eq. (\ref{defk}), it is easily seen that $\mathcal{J}_1^{\pm}(\bm{x}_A, \bm{x}_B)$ can be written as
\begin{equation} \label{IJ1b}
\mathcal{J}^{\pm}_1(\bm{x}_A, \bm{x}_B)=\frac{\sin \psi_{AB}}{1+\cos \psi_{AB}}
(\bm{s}.\bm{n}_A+\bm{s}.\bm{n}_B).
\end{equation}

Obtaining $\mathcal{J}_2^{\pm}$ in a convenient form for discussing the behavior of 
$c\Delta\mathcal{T}_{J_2}^{\pm}(\bm{x}_A,\bm{x}_B)$ when $\psi_{AB}\rightarrow 0$ or 
$\psi_{AB}\rightarrow \pi$ requires some rather tricky transformations detailed in the Appendix. Getting $c\Delta\mathcal{T}_{S}^{\pm}$ is elementary. 

\begin{widetext}
We find:
\begin{eqnarray}
c\Delta\mathcal{T}_{J_1}^{\pm}(\bm{x}_A,\bm{x}_B)=&&-\kappa_1mJ_1\frac{r_0}{b_{\pm}}\frac{\sin \psi_{AB}}{1+\cos \psi_{AB}}(\bm{s}.\bm{n}_A+\bm{s}.\bm{n}_B)   ,\label{TJ1pm} \\
c\Delta\mathcal{T}^{\pm}_{J_2}(\bm{x}_A,\bm{x}_B)=&&\frac{1}{2}\kappa_1mJ_2
\left(\frac{r_0}{b_{\pm}}\right)^2\!\Bigg\lbrace\frac{\sin\psi_{AB}}{1+\cos\psi_{AB}}\left(\frac{b_{\pm}}{r_A}
+\frac{b_{\pm}}{r_B}-\frac{2\kappa_1m}{b_{\pm}}\right) \nonumber \\
&&\times\left[1-2(\bm{s}.\bm{n}_A).(\bm{s}.\bm{n}_B)
-(\bm{s}\times\bm{k}_{AB})^2(1-\cos\psi_{AB})\right]\nonumber \\
&&-\frac{\sin\psi_{AB}}{1+\cos\psi_{AB}}\left[
\left(\frac{b_{\pm}}{r_A}-\frac{\kappa_1m}{b_{\pm}}\right)(\bm{s}.\bm{n}_A)^2
+\left(\frac{b_{\pm}}{r_B}-\frac{\kappa_1m}{b_{\pm}}\right)(\bm{s}.\bm{n}_B)^2\right]
 \nonumber \\
&&+\frac{\kappa_1m}{2b_{\pm}}\Big\lbrack
\left[3(\bm{s}.\bm{k}_{AB})^2-1\right]\psi_{AB}^{\pm}+3\left[(\bm{s}\times\bm{k}_{AB})^2\cos\psi_{AB}
-2(\bm{s}.\bm{n}_A)(\bm{s}.\bm{n}_B)\right]\sin\psi_{AB}\Big\rbrack\Bigg\rbrace, \label{TJ2pm}   \\
c\Delta\mathcal{T}_{S}^{\pm}(\bm{x}_A,\bm{x}_B)=&&-\kappa_1 m \frac{a(\bm{s} .\bm{k}_{AB})}{b_{\pm}} \left[\frac{\sin\psi_{AB}}{1+\cos\psi_{AB}} \left(\frac{b_{\pm}}{r_A}+\frac{b_{\pm}}{r_B}-\frac{2\kappa_1m}{b_{\pm}}\right)
+\frac{\kappa_1m}{b_{\pm}}\psi_{AB}^{\pm}\right]. \label{TS} 
\end{eqnarray}
\end{widetext}

It follows from a result established in Subsect. \ref{S:sss} that the perturbation terms 
given by Eqs. (\ref{TJ1pm})-(\ref{TS}) can be expanded in convergent series in powers of $m$ 
when condition (\ref{cv2}) is satisfied. Considering only the second regime and substituting for 
$1/b_{+}$ from Eq. (\ref{ivbp}), we get for the contributions due to $J_1$ and $J_2$:
\begin{widetext}
\begin{eqnarray}
c\Delta \mathcal{T}_{J_1}^{+}(\bm{x}_A,\bm{x}_B)=&&-\kappa_1mJ_1\frac{r_0R_{AB}}{r_Ar_B}\frac{\bm{s}.\bm{n}_A+\bm{s}.\bm{n}_B}{1+\cos\psi_{AB}}
+O\left(\frac{m^2}{r_{AB}^2}\right),  \label{DTJ1a} \\
c\Delta\mathcal{T}_{J_2}^{+}(\bm{x}_A,\bm{x}_B)=&&\frac{1}{2}\kappa_1mJ_2\frac{r_0^2}{r_Ar_B}\frac{R_{AB}}{1+\cos\psi_{AB}}\bigg\lbrace \frac{1-(\bm{s}.\bm{n}_A)^2}{r_A}
+\frac{1-(\bm{s}.\bm{n}_B)^2}{r_B} \nonumber \\
&&-\left[(\bm{s}\times\bm{k}_{AB})^2(1-\cos\psi_{AB})+2(\bm{s}.\bm{n}_A)(\bm{s}.\bm{n}_B)\right]\left(\!\frac{1}{r_A}+\frac{1}{r_B}\!\right)\!\bigg\rbrace+O\left(\frac{m^2}{r_{AB}^2}\right),\label{DTJ2a} 
\end{eqnarray}
and for the dominant contribution of the angular momentum
\begin{equation} \label{DTSa}
c\Delta\mathcal{T}_{S}^{+}(\bm{x}_A,\bm{x}_B)=-\kappa_1 m a(\bm{s}.\bm{k}_{AB})\frac{r_A+r_B}{r_A r_B}\frac{\sin \psi_{AB}}{1+\cos\psi_{AB}}+O\left(\frac{m^2}{r_{AB}^2}\right).
\end{equation}
\end{widetext}

Equations (\ref{DTJ2a}) and (\ref{DTSa}) are equivalent to formulas previously obtained by other procedures within the first post-Newtonian approximation (see, e.g., \cite{Kopeikin:1997,Kopeikin:2002,Linet:2002,Ciufolini:2003,Leponcin:2008}).

{\it Case where $\bm{s}.\bm{n}_A=\bm{s}.\bm{n}_B=0$}.-- The zeroth-order paths $\Gamma^{+}(\bm{x}_A,\bm{x}_B)$ and $\Gamma^{-}(\bm{x}_A,\bm{x}_B)$ are then confined in the plane passing through $O$ and orthogonal to the axis of rotation of the central body (this plane is the equatorial plane if $O$ is chosen so that $J_1=0$). Equation (\ref{TJ1pm}) reduces to $c\Delta\mathcal{T}_{J_1}^{\pm}(\bm{x}_A,\bm{x}_B)=0$ and Eqs. (\ref{TJ2pm})-(\ref{TS}) are significantly simplified, since $\bm{s}.\bm{n}_A=\bm{s}.\bm{n}_B=0$ implies that $\bm{s}\times\bm{k}_{AB}=0$ and $(\bm{s}.\bm{k}_{AB})^2=1$. More generally, 
the calculation of the integrals $\mathcal{J}_n$ and $\mathcal{K}_n$ is then greatly facilitated since $\bm{s}.\bm{n}(\varphi)=0$ for any $\varphi$, which means that the terms 
$P_n[\bm{s}.\bm{n}(\varphi)]$ and $P'_{n+1}[\bm{s}.\bm{n}(\varphi)]$ reduce to the constants $P_n(0)$ and 
$P'_{n+1}(0)$, respectively. In particular, it follows from the relations 
$P_{2k+1}(0)=P'_{2k}(0)=0$ that 
\begin{equation} \label{TJKod}
c\Delta\mathcal{T}_{J_{2k+1}}^{\pm}(\bm{x}_A,\bm{x}_B)=c\Delta\mathcal{T}_{K_{2k+1}}^{\pm}(\bm{x}_A, \bm{x}_B)=0 
\end{equation}
for $k=0,1,2,\ldots$

%%%%%%%%%%%%%%%%%%%%%%%%%%%%%%%%%%%%%%%%%%%%%%%%%%%%%%%%%%%%%%%%%%%%%%
\subsection{Case where $\psi_{AB}=\pi$} \label{SABo}
%%%%%%%%%%%%%%%%%%%%%%%%%%%%%%%%%%%%%%%%%%%%%%%%%%%%%%%%%%%%%%%%%%%%%%

According to the considerations developed in Subsec. \ref{Samb}, the perturbation terms depend 
now on an arbitrary unit vector $\bm{k}$ orthogonal to $\bm{n}_A$. In this case, $\psi_{AB}^{\pm}=\pm\pi$ and the function $1/r_{\pm}(\varphi)$ occurring in the integrals involved in Eqs. (\ref{TJn}), (\ref{iTS}) and (\ref{iTKn}) is given by Eq. (\ref{trj1}). Taking into account the expression of $b_{\pm}$ supplied by Eq. (\ref{bpmpi}), we get expressions as follow for the terms which must be added to the zeroth-order TTFs given by Eq. (\ref{T0pi}):
\begin{widetext}
\begin{eqnarray}
c\Delta\mathcal{T}^{\pm}_{J_1}(r_A\bm{n}_A,-r_B\bm{n}_A;\bm{k})
=&&\mp J_1 r_0\sqrt{\frac{2\kappa_1m(r_A+r_B)}{r_Ar_B}}
[(\bm{s}\times\bm{k}).\bm{n}_A], \label{lTJ1e} \\
c\Delta\mathcal{T}^{\pm}_{J_2}(r_A\bm{n}_A,-r_B\bm{n}_A;\bm{k})
=&&\frac{1}{2}J_2\frac{r_0^2(r_A+r_B)}{r_Ar_B}\Bigg\lbrace
\sqrt{1+\frac{2\kappa_1m}{r_A+r_B}}\left[2(\bm{s}.\bm{k})^2+(\bm{s}.\bm{n}_A)^2-1\right]  
\nonumber \\   
&&+\sqrt{\frac{\kappa_1m(r_A+r_B)}{2r_Ar_B}}
\left[\pi\frac{3(\bm{s}.\bm{k})^2-1}{4}\mp2\frac{r_B-r_A}{r_A+r_B}
[(\bm{s}\times\bm{k}).\bm{n}_A](\bm{s}.\bm{n}_A)\right]\Bigg\rbrace 
\label{lTJ2e}
\end{eqnarray}
and
\begin{eqnarray}
c\Delta\mathcal{T}^{\pm}_{S}(r_A\bm{n}_A,-r_B\bm{n}_A;\bm{k})
=&&\mp a(\bm{s}.\bm{k})\sqrt{\frac{2\kappa_1m(r_A+r_B)}{r_Ar_B}}\left[\sqrt{1+\frac{2\kappa_1m}{r_A+r_B}} 
+\frac{\pi}{2}\sqrt{\frac{\kappa_1 m(r_A+r_B)}{2r_Ar_B}}\right]. \label{lTSe}
\end{eqnarray}
\end{widetext}

It follows from these formulas that the perturbation terms 
obtained with our procedure are continuous functions when the emitter and the receiver are located in diametrically opposite directions. Indeed, return to Eqs. (\ref{TJ1pm})-(\ref{TS}) and let $\bm{n}_B$ tend to $-\bm{n}_A$ in such a way that $\bm{k}_{AB}\rightarrow\bm{k}$ when $\psi_{AB}\rightarrow \pi$. Using the expansion
\begin{widetext}
\begin{equation} \label{bpml}
b_{\pm}=\pm\sqrt{\frac{2\kappa_1mr_Ar_B}{r_A+r_B}} +\frac{r_Ar_B}{2(r_A+r_B)}\sqrt{1+\frac{2\kappa_1m}{r_A+r_B}}
(\pi-\psi_{AB})+O\left[(\pi-\psi_{AB})^2\right],
\end{equation}
\end{widetext}
it can be checked that Eqs. (\ref{lTJ1e})-(\ref{lTSe}) are recovered.

%%%%%%%%%%%%%%%%%%%%%%%%%%%%%%%%%%%%%%%%%%%%%%%%%%%%%%%%%%%%%%%%%%%%%%%%%%%%%%%%%%%%%%%
\section{Application to the Cassini experiment} \label{sec:num}
%%%%%%%%%%%%%%%%%%%%%%%%%%%%%%%%%%%%%%%%%%%%%%%%%%%%%%%%%%%%%%%%%%%%%%%%%%%%%%%%%%%%%%%

Equation (\ref{dTsp}) has been used to discuss in depth the numerical determination of the post-Newtonian parameter $\gamma$ with an accuracy $\sigma_{\gamma}\approx2.3\times10^{-5}$ from the 
data collected during the Cassini mission (see \cite{Ashby:2010} and Refs. therein). For this determination, the relevant configurations were light rays almost grazing the surface of the 
Sun and corresponding to an emitter and a receiver located in almost diametrically opposed 
positions. So it is of interest to check if the condition of convergence (\ref{cv2}) was satisfied 
or not, and then to compare the numerical values of the travel time of light provided by Eq. (\ref{Tsppm}) and by Eq. (\ref{dTsp}), respectively. 

For the discussion, it is convenient to introduce the dimensionless parameters 
\begin{equation} \label{lalb}
l_A=\frac{r_c}{r_A},\quad l_B=\frac{r_c}{r_B}, 
\quad \epsilon_c=\frac{m}{r_c},
\end{equation}
where $r_c$ is the Euclidean distance between the origin $O$ and the straight line passing 
through $\bm{x}_A$ and $\bm{x}_B$, given by 
\begin{equation} \label{rc}
r_c=\frac{r_Ar_B\sin\psi_{AB}}{R_{AB}}.
\end{equation} 

It is clear that $l_A\ll 1$ and $l_B\ll 1$ in the Cassini mission. Moreover, it may be assumed that $r_c \geq r_{\odot}$, with $r_{\odot}$ being the radius of the Sun ($r_{\odot}$ = 696,000 km), which means that $\epsilon_c\leq2.15\times10^{-6}$. 

Since we are concerned here with configurations such that $\pi/2<\psi_{AB}<\pi$, it is easily inferred from Eqs. (\ref{RAB}) and (\ref{rc}) that 
\begin{equation} \label{cpsi}
\cos\psi_{AB}=l_Al_B-\sqrt{(1-l_A^2)(1-l_B^2)}.
\end{equation} 
Consequently, the necessary and sufficient condition of convergence expressed by Eq. 
(\ref{cv2}) can be written in the form
\begin{eqnarray} 
\sqrt{(1-l_A^2)(1-l_B^2)}\leq &&1+l_Al_B-4\kappa_1\epsilon_c \nonumber \\
&&\times(l_A+l_B-2\kappa_1\epsilon_cl_Al_B).\quad \label{cv4}
\end{eqnarray}
Squaring each side of Eq. (\ref{cv4}) and then simplifying, it is easily checked that a condition like 
\begin{equation} \label{cv5}
l_A+l_B-8\kappa_1\epsilon_c(1+l_Al_B)\geq 0
\end{equation}
is sufficient to insure the convergence of the expansion given by Eqs. (\ref{Tspp})-(\ref{dTsp}).
For the Cassini mission, we have
\begin{equation} \label{rau}
r_A\geq 1 \text{au}, \quad r_B= 1 \text{au},
\end{equation}
where au denotes the astronomical unit. Hence
\begin{equation} \label{lau}
l_A+l_B>4.64\times 10^{-3}
\end{equation} 
since $r_c \geq r_{\odot}$. On the other hand, the inequality $1+l_Al_B\leq2$ is always satisfied since $r_c\leq\text{inf}(r_A,r_B)$. As a consequence, assuming 
that $\kappa_1=2$ leads to
\begin{equation} \label{lau1}
8\kappa_1\epsilon_c(1+l_Al_B)<6.86\times10^{-5}
\end{equation} 
for any light ray travelling in the Solar System. It follows from Eqs. (\ref{lau}) and 
(\ref{lau1}) that condition (\ref{cv5}) was fully satisfied in the part of the Cassini 
mission exploited for determining $\gamma$. The TTF expansion whose first terms are given 
by Eq. (\ref{dTsp}) was therefore rightfully used.

To numerically compare the results provided by the functions $c\mathcal{T}_{sph}^{+}$ and $c\mathcal{T}_{sph}^{\ast}$, it is necessary to know the expressions of $\sin\psi_{AB}$ and 
$R_{AB}$. Equation (\ref{cpsi}) yields  
\begin{equation} \label{spsi}
\sin\psi_{AB}=l_A\sqrt{1-l_B^2}+l_B\sqrt{1-l_A^2}.
\end{equation}
The expression of $R_{AB}$ is then straightforwardly obtained by substituting for 
$\sin\psi_{AB}$ from Eq. (\ref{spsi}) into Eq. (\ref{rc}):
\begin{equation} \label{RAB2}
R_{AB}=r_c\frac{l_A\sqrt{1-l_B^2}+l_B\sqrt{1-l_A^2}}{l_Al_B}.
\end{equation}
Furthermore, it is convenient to use the ratio $m/b_{+}$ written in the following form, directly deduced from Eq. (\ref{bpm}):
\begin{widetext}
\begin{equation} \label{imb}
\frac{m}{b_{+}}=\frac{\sqrt{1+\cos\psi_{AB} 
+2\kappa_1\epsilon_c\left(l_A+l_A+\sin\psi_{AB}\right)}-\sqrt{1+\cos\psi_{AB} 
+2\kappa_1\epsilon_c\left(l_A+l_A-\sin\psi_{AB}\right)}}{2\kappa_1\sqrt{1-\cos\psi_{AB}}},
\end{equation}
\end{widetext}
where $\cos\psi_{AB}$ and $\sin\psi_{AB}$ have to be replaced by their expressions yielded 
by Eq. (\ref{cpsi}) and Eq. (\ref{spsi}), respectively.

The quantities $c\mathcal{T}^{+}_{sph}$ and $c\mathcal{T}^{\ast}_{sph}$ can be expressed
as functions of $l_A, l_B, r_c, m,\kappa_1,\kappa_2,\kappa_3$. For the comparison, we assume that the parameters $\kappa_1,\kappa_2$ and $\kappa_3$ take their values predicted by general relativity, namely
\begin{equation} \label{k123}
\kappa_1=2,\quad \kappa_2=\frac{15}{4},\quad \kappa_3=\frac{9}{2}.
\end{equation} 

Given $r_A$ and $r_B$, $\vert c\mathcal{T}^{+}_{sph}-c\mathcal{T}^{\ast}_{sph}\vert$ is a 
decreasing function of $r_c$. So the most favorable estimate will be obtained for $r_c=r_{\odot}$. Then, with $r_B=1$ au, it is easily checked that $\vert c\mathcal{T}^{+}_{sph}-c\mathcal{T}^{\ast}_{sph}\vert$ increases up to 27.35 $\mu$m as $r_A$ increases up to 40 au. 
Our correction is thus much smaller than the uncertainty on the position of the Cassini 
spacecraft, which was typically about 30 cm. 

%%%%%%%%%%%%%%%%%%%%%%%%%%%%%%%%%%%%%%%%%%%%%%%%%%%%%%%%%%%%%%%%%%%%%%%%%%%%%%%%%%%%%%%
\section{Conclusion} \label{sec:concl}
%%%%%%%%%%%%%%%%%%%%%%%%%%%%%%%%%%%%%%%%%%%%%%%%%%%%%%%%%%%%%%%%%%%%%%%%%%%%%%%%%%%%%%%

In this paper we develop a perturbation method allowing to determine a class of TTFs for a 
stationary metric which is a weak perturbation of a background metric being itself stationary. 
This method works when the light rays of the background metric are explicitly known as 
parametric functions only involving the spatial positions of the emitter and the receiver. Equations (\ref{dI1}), (\ref{T1a}) and (\ref{dl}) provide the basic results as simple integrals taken along the zeroth-order light rays.  

When the metric is a stationary perturbation of a Schwarzschild-like metric, the expressions of 
the TTFs in the generic case are given by Eqs. (\ref{DTpm}) and (\ref{T0pm}), the zeroth-order light rays being Keplerian hyperbolas.   

For an isolated, slowly rotating axisymmetric body, the integrals giving the perturbation terms 
due to the mass and spin multipoles can be calculated with any symbolic computer program. We get explicit expressions of these integrals for the mass multipoles $J_1$, $J_2$ and for the 
intrinsic angular momentum $\bm{S}$ of the rotating body. 

The procedure set in the present work carries out a noticeable improvement on the previous 
attempts to determine the time transfer functions since the expressions obtained 
for the TTFs are totally devoid of enhanced terms whatever the positions of the emitter and the receiver. Furthermore, a criterion of convergence of the expansion of the TTFs in series of 
powers of $m$ is supplied by Eqs. (\ref{cv2}) and (\ref{psm}). 

It can be underlined that our calculations remain valid when the emitter and the receiver are in diametrically opposite positions. So the present methodology will certainly be useful for 
modeling the difference in light travel time between two different images occuring in a gravitational lensing configuration. 

%%%%%%%%%%%%%%%%%%%%%%%
\begin{acknowledgments}

This paper is dedicated to the memory of Bernard Linet, with whom we had very stimulating discussions which are at the origin of this work. 

We warmly thank Jean Teyssandier for his careful reading of the manuscript.
\end{acknowledgments}
%%%%%%%%%%%%%%%%%%%%%

%%%%%%%%%%%%%%%%%%%%%%%%%%%
\appendix*
%%%%%%%%%%%%%%%%%%%%%%%%%%%

\begin{widetext}

%%%%%%%%%%%%%%%%%%%%%%%%%%%%%%%%%%%%%%%%%%%%%%%%%%%%%%%%%%%%%%%%%%%%%%%%%%%%%%%%%%%
\section{Calculation of $c\Delta\mathcal{T}_{J_2}^{\pm}$} \label{sec:cTJ2}
%%%%%%%%%%%%%%%%%%%%%%%%%%%%%%%%%%%%%%%%%%%%%%%%%%%%%%%%%%%%%%%%%%%%%%%%%%%%%%%%%%%

Let us put
\begin{equation} \label{Apm}
A^{\pm}(\bm{x}_A, \bm{x}_B)=\int_{0}^{\psi_{AB}^{\pm}}\left\{\sin^2\psi_{AB}-3\left[(\bm{s} .\bm{n}_A)
\sin(\psi_{AB}-\varphi)+(\bm{s} .\bm{n}_B)\sin\varphi\right]^2  \right\}d\varphi
\end{equation}
and
\begin{eqnarray} \label{Bpm}
B^{\pm}(\bm{x}_A, \bm{x}_B;p)&=& \int_{0}^{\psi_{AB}^{\pm}}\left\{\sin^2\psi_{AB}-3\left[(\bm{s} .\bm{n}_A)
\sin(\psi_{AB}-\varphi)+(\bm{s}.\bm{n}_B)\sin\varphi\right]^2  \right\}  \nonumber \\
&& \times\left[\left(\frac{p}{r_A}-1\right)\sin(\psi_{AB}-\varphi)
+\left(\frac{p}{r_B}-1\right)\sin\varphi\right]d\varphi.
\end{eqnarray}

Since $P_2(x)=(3x^2-1)/2$, Eq. (\ref{TJn}) may be written as follows for $n=2$ :
\begin{equation} \label{J2pm0}
c\Delta\mathcal{T}_{J_2}^{\pm}(\bm{x}_A, \bm{x}_B)=\frac{1}{2}\kappa_1mJ_2\left(\frac{r_0}{b_{\pm}}\right)^2\frac{\kappa_1 m}{b_{\pm}}\frac{A^{\pm}(\bm{x}_A, \bm{x}_B)\sin\psi_{AB}+B^{\pm}(\bm{x}_A, \bm{x}_B,p_{\pm})}{\sin^3\psi_{AB}}.
\end{equation}

The problem is reduced to the calculation of $A^{+}(\bm{x}_A, \bm{x}_B)$ and $B^{+}(\bm{x}_A, \bm{x}_B;p)$ since it is easily checked that 
\begin{equation} \label{ABmp}
\frac{A^{-}(\bm{x}_A,\bm{x}_B)\sin\psi_{AB}+B^{-}(\bm{x}_A,\bm{x}_B;p)}{\sin^3\psi_{AB}}=\frac{A^{+}(\bm{x}_A,\bm{x}_B)\sin\psi_{AB}+B^{+}(\bm{x}_A,\bm{x}_B;p)}{\sin^3\psi_{AB}}-2\pi P_2(\bm{s}.\bm{k}_{AB}).
\end{equation}

Using a symbolic computer program and grouping the terms involving the parameter $p$ give
\begin{eqnarray}
\frac{A^{+}(\bm{x}_A, \bm{x}_B)\sin\psi_{AB}+B^{+}(\bm{x}_A,\bm{x}_B;p)}{\sin^3\psi_{AB}}&=&
\left[1-3\frac{(\bm{s}.\bm{n}_A)^2+(\bm{s}.\bm{n}_B)^2-2(\bm{s}.\bm{n}_A)(\bm{s}.\bm{n}_B)\cos\psi_{AB}}{2\sin^2\psi_{AB}}\right]\psi_{AB}  \nonumber \\
&& +3\frac{[(\bm{s}.\bm{n}_A)^2+(\bm{s}.\bm{n}_B)^2]\cos\psi_{AB}-2(\bm{s}.\bm{n}_A)(\bm{s}.\bm{n}_B)}{2\sin\psi_{AB}} \nonumber \\
&& +\frac{\sin\psi_{AB}}{1+\cos\psi_{AB}}p\bigg\lbrack\frac{1-(\bm{s}.\bm{n}_A)^2}{r_A}+\frac{1-(\bm{s}. \bm{n}_B)^2}{r_B}\nonumber \\
&&-\frac{(\bm{s}.\bm{n}_A+\bm{s}.\bm{n}_B)^2}{1+\cos\psi_{AB}}\left(\!\frac{1}{r_A}+\frac{1}{r_B}\!\right)\bigg\rbrack  \nonumber \\
&&+\frac{\sin\psi_{AB}}{1+\cos\psi_{AB}}\left[(\bm{s}.\bm{n}_A)^2+(\bm{s}.\bm{n}_B)^2-2+2\frac{(\bm{s}.\bm{n}_A+\bm{s}.\bm{n}_B)^2}{1+\cos\psi_{AB}}\right]. \label{ABsp}
\end{eqnarray}

Several terms involved in the right-hand side of Eq. (\ref{ABsp}) can be transformed in such a 
way that their apparent divergence for $\psi_{AB}\rightarrow 0$ or $\psi_{AB}\rightarrow \pi$ 
is eliminated, as it is shown in what follows.

1.-- Using the 
identity
\begin{equation} \label{id1a}
(\bm{s}\times(\bm{n}_A\times\bm{n}_B))^2=(\bm{s}.\bm{n}_A)^2+(\bm{s}.\bm{n}_B)^2-2(\bm{s}.\bm{n}_A)(\bm{s}.\bm{n}_B)\cos\psi_{AB},
\end{equation}
and then taking into account Eq. (\ref{defk}), it is easily seen that 
\begin{equation} \label{id1b}
\frac{(\bm{s}.\bm{n}_A)^2+(\bm{s}.\bm{n}_B)^2-2(\bm{s}.\bm{n}_A)(\bm{s}.\bm{n}_B)\cos\psi_{AB}}{\sin^2\psi_{AB}}=(\bm{s}\times\bm{k}_{AB})^2
=1-(\bm{s}.\bm{k}_{AB})^2.
\end{equation}
As a consequence
\begin{equation} \label{id1c}
1-3\frac{(\bm{s}.\bm{n}_A)^2+(\bm{s}.\bm{n}_B)^2-2(\bm{s}.\bm{n}_A)(\bm{s}.\bm{n}_B)\cos\psi_{AB}}{2\sin^2\psi_{AB}}=
\frac{3(\bm{s}.\bm{k}_{AB})^2-1}{2}=P_2(\bm{s}.\bm{k}_{AB}).
\end{equation} 

2.-- From Eq. (\ref{id1b}), it is inferred that
\begin{equation} \nonumber
\frac{[(\bm{s}.\bm{n}_A)^2+(\bm{s}.\bm{n}_B)^2]\cos\psi_{AB}-2(\bm{s}.\bm{n}_A)(\bm{s}.\bm{n}_B)(1-\sin^2\psi_{AB})}{\sin\psi_{AB}}=[1-(\bm{s}.\bm{k}_{AB})^2]\sin\psi_{AB}\cos\psi_{AB}.
\end{equation}
Therefore
\begin{equation} \label{id2}
\frac{[(\bm{s}.\bm{n}_A)^2+(\bm{s}.\bm{n}_B)^2]\cos\psi_{AB}-2(\bm{s}.\bm{n}_A)(\bm{s}.\bm{n}_B)}{\sin\psi_{AB}}=
\left\{[1-(\bm{s}.\bm{k}_{AB})^2]\cos\psi_{AB}-2(\bm{s}.\bm{n}_A)(\bm{s}.\bm{n}_B)\right\}\sin\psi_{AB}.
\end{equation}

3.-- Equation (\ref{id1a}) may be rewritten in the form
\begin{equation} \label{id3a}
(\bm{s}.\bm{n}_A)^2+(\bm{s}.\bm{n}_B)^2=(\bm{s}\times\bm{k}_{AB})^2\sin^2\psi_{AB}+2(\bm{s}.\bm{n}_A)(\bm{s}.\bm{n}_B)\cos\psi_{AB}.
\end{equation}
This relation implies
\begin{equation} \label{id3b}
[(\bm{s}.\bm{n}_A)+(\bm{s}.\bm{n}_B)]^2=\left[(\bm{s}\times\bm{k}_{AB})^2(1-\cos^2\psi_{AB})+2(\bm{s}.\bm{n}_A)(\bm{s}.\bm{n}_B)\right](1+\cos\psi_{AB}).
\end{equation}
As a consequence
\begin{equation} \label{id3c}
\frac{[(\bm{s}.\bm{n}_A)+(\bm{s}.\bm{n}_B)]^2}{1+\cos\psi_{AB}}=\left[1-(\bm{s}.\bm{k}_{AB})^2\right](1-\cos\psi_{AB})+2(\bm{s}.\bm{n}_A)(\bm{s}.\bm{n}_B).
\end{equation}

Inserting Eqs. (\ref{id1c}), (\ref{id2}) and (\ref{id3c}) into Eq. (\ref{ABsp}), and then taking Eq. (\ref{ABmp}) into account, we get 
\begin{eqnarray}
\frac{A^{+}(\bm{x}_A, \bm{x}_B)\sin\psi_{AB}+B^{+}(\bm{x}_A,\bm{x}_B;p)}{\sin^3\psi_{AB}}&=&
P_2(\bm{s}.\bm{k}_{AB})\psi_{AB}\nonumber \\
&&+\frac{3}{2}\left\{\left[1-(\bm{s}.\bm{k}_{AB})^2\right]\cos\psi_{AB}-2(\bm{s}.\bm{n}_A)(\bm{s}.\bm{n}_B)\right\}\sin\psi_{AB}   \nonumber \\
&&+\frac{\sin\psi_{AB}}{1+\cos\psi_{AB}}\big\lbrack(\bm{s}.\bm{n}_A)^2+(\bm{s}.\bm{n}_B)^2 +4(\bm{s}.\bm{n}_A)(\bm{s}.\bm{n}_B) \nonumber \\
&&-2\cos\psi_{AB}-2(\bm{s}.\bm{k}_{AB})^2(1-\cos\psi_{AB})\big\rbrack\nonumber \\
&& +\frac{\sin\psi_{AB}}{1+\cos\psi_{AB}}p\left[\frac{1-(\bm{s}.\bm{n}_A)^2}{r_A}+\frac{1-(\bm{s}. \bm{n}_B)^2}{r_B} \right. \nonumber \\
&&\left. -\left\{[1-(\bm{s}.\bm{k}_{AB})^2](1-\cos\psi_{AB})+2(\bm{s}.\bm{n}_A)(\bm{s}.\bm{n}_B)\right\}\frac{r_A+r_B}{r_Ar_B}\right]. \label{ABsp1}
\end{eqnarray}

Hence the formula (\ref{TJ2pm}) when Eq. (\ref{ppm}) is taken into account and the terms are 
suitably grouped together.

\end{widetext}

\nocite{*}

\bibliography{apssamp}% Produces the bibliography via BibTeX.

\end{document}